\documentclass[aip,jcp,preprint,amsmath,amssymb,nofootinbib]{revtex4-2}
\usepackage{graphicx}
\usepackage{bm}
\usepackage{braket}
\usepackage{tikz}
\usetikzlibrary{decorations.markings,arrows.meta,positioning,calc,shapes.misc}
\usepackage[colorlinks=true,allcolors=blue]{hyperref}

\newcommand{\Hbar}{\bar{H}}
\newcommand{\vtx}[1]{\Gamma^{#1}}

\tikzset{
  fermion/.style={thick,postaction={decorate},decoration={markings,mark=at position 0.55 with {\arrow{Stealth[length=5pt]}}}},
  interaction/.style={thick,dashed},
  vertexdot/.style={circle,fill,inner sep=1.3pt},
  respbox/.style={draw,thick,fill=gray!15,minimum width=1.5cm,minimum height=0.9cm,rounded corners=2pt},
}

\begin{document}

\title{A response-based self-energy from equation-of-motion coupled-cluster theory\\
and its reduction to GW, GW+SOSEX, and GF2}

\author{Ajith Perera}
\email{perera.ajith@gmail.com}
\affiliation{Quantum Theory Project, University of Florida, Gainesville, Florida 32611, USA}
\author{Pierre-Fran\c{c}ois Loos}
\affiliation{Univ Toulouse, CNRS, Laboratoire de Chimie et Physique Quantiques, Toulouse, France}

\date{\today}

\begin{abstract}
We construct a frequency-dependent one-particle correlation self-energy by coupling a reference
Green's function to the neutral excitation response obtained from equation-of-motion
coupled-cluster theory with singles and doubles (EOM-CCSD). Direct, mixed
direct--antisymmetrized, and fully antisymmetrized choices of the external interaction vertices
define three approximations, all admitting an explicit spectral representation without numerical
frequency integration. In the bare-state limit, the mixed and fully antisymmetrized forms recover
second-order Green's-function theory (GF2), whereas the direct form retains only its ring
contribution. With a direct random-phase approximation (RPA) response, the direct and mixed forms
reduce to $G_0W_0$ and $G_0W_0$ plus second-order screened exchange (SOSEX), respectively. For
molecular ionization potentials, replacing the RPA response by EOM-CCSD generally lowers the
ionization energies, whereas introducing exchange at one external vertex raises them. These
substantial opposing effects partially compensate, although the mixed approximation does not
systematically improve upon conventional $G_0W_0$, either against IP-EOM-CCSD or QUESTDB
reference energies (aug-cc-pVDZ mean absolute errors relative to the QUESTDB theoretical best
estimates: 0.45 vs 0.36~eV). A third-order perturbative analysis reveals that the three
approximations retain different contributions and weights of particle--hole diagrams while
omitting explicit particle--particle and hole--hole ladder contributions and the static
self-energy. In particular, the fully antisymmetrized form recovers GF2 at second order but
assigns incorrect weights to its retained third-order diagrams. These findings highlight the
importance of a balanced treatment of screening and vertex corrections in constructing accurate
self-energy approximations.
\end{abstract}

\maketitle

\section{Introduction}

The $GW$ approximation\cite{hedin1965} is the standard many-body method for charged
excitations of solids and, increasingly, of molecules.\cite{reining2018,golze2019} In its
one-shot form on a Hartree--Fock or Kohn--Sham reference, $G_0W_0$, it gives valence
ionization potentials (IPs) of molecules to a few tenths of an electronvolt,\cite{vansetten2015}
at a cost far below that of wave-function methods of similar accuracy. Formally, $GW$ is
the approximation obtained from Hedin's equations by setting the vertex $\Gamma$ to unity in
both the polarizability and the self-energy. The screened interaction $W$ is then built from the
direct random-phase approximation (RPA) polarizability, and the self-energy contains no
exchange beyond first order.

Going beyond $GW$ by adding vertex corrections has proved surprisingly difficult.
Vertex corrections can be added to the polarizability, to the self-energy, or to both;
they have been studied for solids\cite{shishkin2007,gruneis2014} and
molecules,\cite{romaniello2009,ren2015,kuwahara2016,maggio2017,wang2021,forster2022,bruneval2024}
and the conclusion of most of these studies is the same: improving one ingredient alone
often makes the results worse. Lewis and Berkelbach showed that replacing the RPA
polarizability by a coupled-cluster one does not improve molecular IPs.\cite{lewis2019}
Marie and Loos, working within parquet theory, attribute the accuracy of $GW$ to ``a
fortuitous cancellation of vertex corrections in the polarizability and self-energy'' and
argue that the two-body scattering channels must be treated on an equal
footing.\cite{marie2025,bickers1989} F\"orster and Bruneval quantified this cancellation for
molecules with a range of non-local vertices, at first order and summed to all orders, built
from bare and statically screened interactions and placed in both the polarizability and the
self-energy: the two vertices essentially compensate, and only a consistent treatment of both
gives realistic quasiparticle energies.\cite{forster2024} Marie, T\"olle and Loos recast the
second-order vertex-corrected (G3W2) self-energy in the algebraic-diagrammatic construction
(ADC), which enforces the sum-over-states form and positive semi-definiteness of the exact
self-energy and connects vertex corrections built on $W$ with the ADC hierarchy built on the
bare interaction.\cite{marie2026adc}

Coupled-cluster (CC) theory\cite{bartlett2007,shavitt2009} approaches the same quantities
from the other side. Equation-of-motion CC (EOM-CC) gives ionized and excited states
directly,\cite{stanton1993,stanton1994} and the CC Green's function was formulated early
on.\cite{nooijen1992} The connections between the two families are now well
documented: RPA is ring-CCD,\cite{scuseria2008} RPA excitation energies follow from an
approximate EOM-CCD,\cite{berkelbach2018,rishi2020} $GW$ can be written as an
EOM-like eigenvalue problem,\cite{lange2018,bintrim2021} exact relationships between $GW$
and EOM-CC have been derived,\cite{quintero2022,tolle2023} and a diagrammatic definition of
the irreducible CC self-energy has recently been given.\cite{coveney2025,tolle2026} These
works start from the self-energy or from the CC equations and ask which diagrams they
contain.

Here we take a third route, one that comes from the theory of NMR spin--spin coupling
constants.\cite{perera1996,perera2006} A coupling constant is a second-order property: two
one-electron operators, one at each nucleus, joined by the linear response of the
electrons. The self-energy has the same structure once one of the two orbital lines of
each two-electron interaction is held fixed: the interaction then acts as a one-body
perturbation on the remaining electrons, and two such perturbations are joined by their
response. With the EOM-CCSD response function this gives a self-energy whose screening
is correlated at the CC level and whose two vertices can be attached in direct, mixed or
antisymmetrized form. Evaluating it over all eigenstates of the EOM-CCSD
similarity-transformed Hamiltonian, obtained by complete diagonalization, makes it a sum
over states: for the given states the self-energy is a rational function known on the
whole real axis, the quasiparticle equation carries no frequency-integration or
analytic-continuation error, and satellites come out with the main lines.

The purpose of this paper is to document this construction, to show how it reduces to known
theories, and to use it to separate the effects of the screening and of the self-energy vertex.
With bare orbital-difference states the construction reproduces second-order Green's-function
theory (GF2),\cite{szabo1989} and with direct-RPA states its direct and mixed forms reduce to
$G_0W_0$@HF and $G_0W_0$+SOSEX.\cite{ren2015} With EOM-CCSD states the same code gives
$GW$ with a CC polarizability, the same with an exchange vertex in the self-energy, and the
form with both vertices antisymmetrized. Screening (RPA or CCSD) and self-energy vertex
(direct or with exchange) can thus be changed one at a time, against one reference.
For eight first-row systems the two changes turn out to be nearly additive and of opposite sign,
the compensation between vertex corrections discussed by Marie and Loos.\cite{marie2025}

A second aim is to identify the self-energy diagrams that the three forms contain, rather than to
assign them $GW$, SOSEX or ladder labels by analogy. Expanding the CC amplitudes, the EOM Jacobian
and the transition densities in the fluctuation potential (Sec.~\ref{sec:pt} and
Appendices~\ref{app:resp} and~\ref{app:sigma3}), the zeroth-order response gives the
second-order self-energy and the first-order response gives all third-order terms. This separates
the correlation inside the response from the exchange introduced at the two external vertices,
and shows which third-order diagrams each form contains, with which weight, and which it lacks.
The weights are derived algebraically.

The paper is organized as follows. Section~\ref{sec:theory} develops the construction, its
reductions and its third-order structure in antisymmetrized diagrammatic form.
Section~\ref{sec:comp} gives the computational details, Sec.~\ref{sec:results} the results, and
Sec.~\ref{sec:concl} the conclusions. The perturbative expansion of the response and the
third-order expressions are given in the appendices.

\section{Theory}\label{sec:theory}

We use spin orbitals $p,q,r,s,\dots$ of a Hartree--Fock reference $\ket{0}$ with orbital
energies $\epsilon_p$; $i,j$ label occupied and $a,b$ virtual orbitals, and $m$ is an
orbital of either kind; $n$ labels the neutral excited states that carry the response
(EOM-CCSD, direct-RPA or bare orbital-difference states). Two-electron integrals are $\braket{pq|rs}=(pr|qs)$ and
$\braket{pq||rs}=\braket{pq|rs}-\braket{pq|sr}$, and $E_{rs}=a_r^\dagger a_s$.

\subsection{Coupling of two one-body perturbations through the response}

The response contribution to the reduced NMR spin--spin coupling between nuclei $A$ and $B$ is the
linear-response function
of two one-electron operators $h_A=\sum_{rs}h^A_{rs}E_{rs}$ and $h_B$,
\begin{equation}
K_{AB}\;\propto\;\langle\!\langle h_A;h_B\rangle\!\rangle_{\omega=0},
\label{eq:Kab}
\end{equation}
where for one-body operators $X$ and $Y$
\begin{equation}
\langle\!\langle X;Y\rangle\!\rangle_z=\sum_{n}\left[\frac{X^R_n\,Y^L_n}{z-\omega_n}
-\frac{Y^R_n\,X^L_n}{z+\omega_n}\right].
\label{eq:chiXY}
\end{equation}
Here $T=T_1+T_2$ solves the CCSD equations, $\Hbar=e^{-T}He^{T}$, $E_\mathrm{CC}=\bra{0}\Hbar\ket{0}$,
and $\Lambda=\Lambda_1+\Lambda_2$ solves the ground-state left equations
$\bra{0}(1+\Lambda)(\Hbar-E_\mathrm{CC})=0$. In EOM-EE-CCSD\cite{stanton1993} the
excitation energies and right and left states satisfy
$(\Hbar-E_\mathrm{CC})R_n\ket{0}=\omega_n R_n\ket{0}$ and
$\bra{0}L_n(\Hbar-E_\mathrm{CC})=\omega_n\bra{0}L_n$, with $R_n$ and $L_n$ in the space of
single and double excitations (de-excitations for $L_n$), biorthonormal,
$\bra{0}L_nR_{n'}\ket{0}=\delta_{nn'}$, and $\bra{0}L_n\ket{0}=0$. The transition moments are
\begin{align}
X^R_n&=\sum_{rs}X_{rs}\,\gamma^R_n(rs),&\gamma^R_n(rs)&=\bra{0}(1+\Lambda)\bar E_{rs}R_n\ket{0},
\label{eq:gammaR}\\
X^L_n&=\sum_{rs}X_{rs}\,\gamma^L_n(rs),&\gamma^L_n(rs)&=\bra{0}L_n\bar E_{rs}\ket{0},
\label{eq:gammaL}
\end{align}
with $\bar E_{rs}=e^{-T}E_{rs}e^{T}$ and $\Lambda$ the ground-state de-excitation operator. In
Eq.~\eqref{eq:gammaR} $R_n$ includes its reference component $r_{0,n}$, fixed by requiring
$R_n\ket{0}$ to be a right eigenvector of $\Hbar-E_\mathrm{CC}$ in the space that includes
$\ket{0}$, so that $\bra{0}(1+\Lambda)R_n\ket{0}=0$. Equation~\eqref{eq:gammaR} is an operator product, not
the commutator $\bra{0}(1+\Lambda)[\bar E_{rs},R_n]\ket{0}$; the two differ already at first
order in the fluctuation potential (Appendix~\ref{app:resp}). This is the EOM-CC form of the response function used for spin--spin
couplings;\cite{perera1996} the sum over $n$ is the spectral representation of the resolvent
$(z-\bm{J})^{-1}$, where $J_{\mu\nu}=\bra{\mu}\Hbar-E_\mathrm{CC}\ket{\nu}$ is the EOM-CCSD Jacobian in the space of excited determinants $\ket{\mu}$. Its relation to CC linear-response (LR) theory is as follows. A static response property such as
$K_{AB}$ is a mixed second derivative of the energy with respect to two perturbation strengths.
As in CC-LR theory, the reference determinant is held fixed (orbital-unrelaxed response), and
the operators $h_A$, $h_B$, and $h^{(pm)}$ below, do not depend on the basis set, so no
basis-function derivatives arise. The fixed-reference analytic CC second
derivative\cite{salter1989a,salter1989b} then has three terms: two in which one perturbation
enters through $\Hbar$ and the other through the first-order amplitudes, weighted by
$(1+\Lambda)$, and one quadratic in the first-order amplitudes. With all three it is identical
to the CC-LR function obtained from the time-dependent CC
Lagrangian.\cite{koch1990,christiansen1998} Equation~\eqref{eq:chiXY} is instead the EOM-CC
sum-over-states form.\cite{stanton1993pol,perera1996} In truncated CC it is not identical to
CC-LR: the two differ by terms involving the perturbed ground-state amplitudes, and their
transition moments differ,\cite{kallay2004} and they coincide only in the full-CI limit. We use
the EOM-CC form throughout because it gives the response explicitly as a sum over states at
every frequency, which is what a self-energy requires.
Because $\Hbar$ is not Hermitian, $\gamma^L_n$ is not the conjugate of $\gamma^R_n$, and
the products $X^R_nY^L_n$ need not be positive. The coupling is thus two
perturbations, each written as a transition density contracted with an operator, joined by
a propagator of $\Hbar$ [Fig.~\ref{fig:diagrams}(c)].

\subsection{The two-electron interaction as a one-body perturbation}

Consider an electron in orbital $p$ that interacts with the other electrons, leaves in an
intermediate orbital $m$, interacts again, and leaves in $q$. Holding the pair $(p,m)$
fixed, the two-electron interaction acts on the remaining electrons as the one-body
operator
\begin{equation}
h^{(pm)}=\sum_{rs}\braket{pr||ms}\,E_{rs}.
\label{eq:hpm}
\end{equation}
This plays the role of $h_A$ in Eq.~\eqref{eq:Kab}. Its right and left transition moments are the antisymmetrized vertices
\begin{equation}
\vtx{R}_n(p,m)=\sum_{rs}\braket{pr||ms}\,\gamma^R_n(rs),\qquad
\vtx{L}_n(p,m)=\sum_{rs}\braket{pr||ms}\,\gamma^L_n(rs),
\label{eq:vertices}
\end{equation}
whose direct and exchange parts we write as $\vtx{}=V-X$, with
$V_n(p,m)=\sum_{rs}\braket{pr|ms}\gamma_n(rs)=\sum_{rs}(pm|rs)\gamma_n(rs)$ and
$X_n(p,m)=\sum_{rs}\braket{pr|sm}\gamma_n(rs)=\sum_{rs}(ps|rm)\gamma_n(rs)$ for either density.
In $V$ the external pair $(p,m)$ forms one charge distribution and the transition density the
other, as in the bubble of $GW$; in $X$ the external line is exchanged into the response. For a
closed-shell reference and real orbitals $(pm|rs)=(pm|sr)$, so $V$ depends only on the spin-summed, symmetric part of the
transition density: spin-triplet states do not enter $V$, but they do enter $X$.

\subsection{Self-energy}

Joining two such vertices by the response of the other electrons and propagating the
intermediate orbital $m$ with the reference Green's function gives the correlation
self-energy, a frequency convolution of the time-ordered reference propagator
$G^0_m(\omega)=[\omega-\epsilon_m\pm i0^+]^{-1}$ ($+$ for $m$ virtual, $-$ for $m$ occupied)
with the response. Closing the contour over the poles of $G^0$ and of the response, the term
with a hole as intermediate line (removal) picks up the forward part of Eq.~\eqref{eq:chiXY} and
the term with a particle (addition) its backward part, each evaluated at a frequency shifted by
the orbital energy of the intermediate line (Appendix~\ref{app:resp}):
\begin{equation}
\Sigma_{pq}(\omega)=\kappa\sum_n\left[\sum_i\frac{P^R_n(i,p)\,Q^L_n(q,i)}{\omega-\epsilon_i+\omega_n}
+\sum_a\frac{Q^R_n(q,a)\,P^L_n(a,p)}{\omega-\epsilon_a-\omega_n}\right],
\label{eq:sigma}
\end{equation}
where $P$ is the vertex at the end of the incoming line $p$ and $Q$ the vertex at the end of
the outgoing line $q$, each either $V$ or $\vtx{}$ of Eq.~\eqref{eq:vertices}. The first term
has the intermediate line as a hole ($m=i$, the two-hole--one-particle time ordering), the
second as a particle ($m=a$, two-particle--one-hole). The left moment creates the state and the
right moment removes it; in the hole ordering the state is created at the $q$ vertex, in the
particle ordering at the $p$ vertex, which is why the right and left moments exchange places
between the two terms. Equation~\eqref{eq:sigma}
is the time-ordered self-energy with the infinitesimals suppressed: the first term carries
$-i0^+$ and the second $+i0^+$ in its denominator (the retarded function has $+i0^+$ in
both). Only real $\omega$ away from the poles is used below. The conventional frequency-convolution derivation assumes a neutral response with the appropriate time-ordered analytic structure. For complex EOM-CCSD excitation energies, Eq.~\eqref{eq:sigma} is used as the algebraic definition of the approximate self-energy, without implying that the resulting function is causal. In the
language of Eq.~\eqref{eq:chiXY} it is
\begin{equation}
\Sigma_{pq}(\omega)=-\kappa\sum_i\langle\!\langle h^{(ip)};h^{(qi)}\rangle\!\rangle^{+}_{\epsilon_i-\omega}
+\kappa\sum_a\langle\!\langle h^{(qa)};h^{(ap)}\rangle\!\rangle^{+}_{\omega-\epsilon_a},
\label{eq:sigma_chi}
\end{equation}
where $^{+}$ denotes the forward (first) term of Eq.~\eqref{eq:chiXY}; in the second sum the
backward part of the response has been written in this form with the two operators exchanged.
Here $h^{(pm)}$ is
Eq.~\eqref{eq:hpm} or its direct part, according to the vertex. The self-energy is thus a sum of
spin--spin-coupling-like propagators between pairs of effective one-body operators. The
prefactor and signs of Eq.~\eqref{eq:sigma} are fixed by the bare-state limit below, where they
are shown term by term to give GF2.
In short, the self-energy follows in three steps: fixing the outer line $m\to p$ turns the
interaction into the one-body operator $h^{(pm)}$ [Eq.~\eqref{eq:hpm}]; its transition moments
with the EOM-CC states are the vertices [Eq.~\eqref{eq:vertices}]; and the forward (removal) and backward (addition) parts of the response
function, evaluated at frequencies shifted by $\epsilon_m$ and summed over $m$, give
Eq.~\eqref{eq:sigma}.
For a given set of states and vertices Eq.~\eqref{eq:sigma} involves no further
approximation and no frequency integration, so $\Sigma(\omega)$ is a rational function
known everywhere on the real axis.

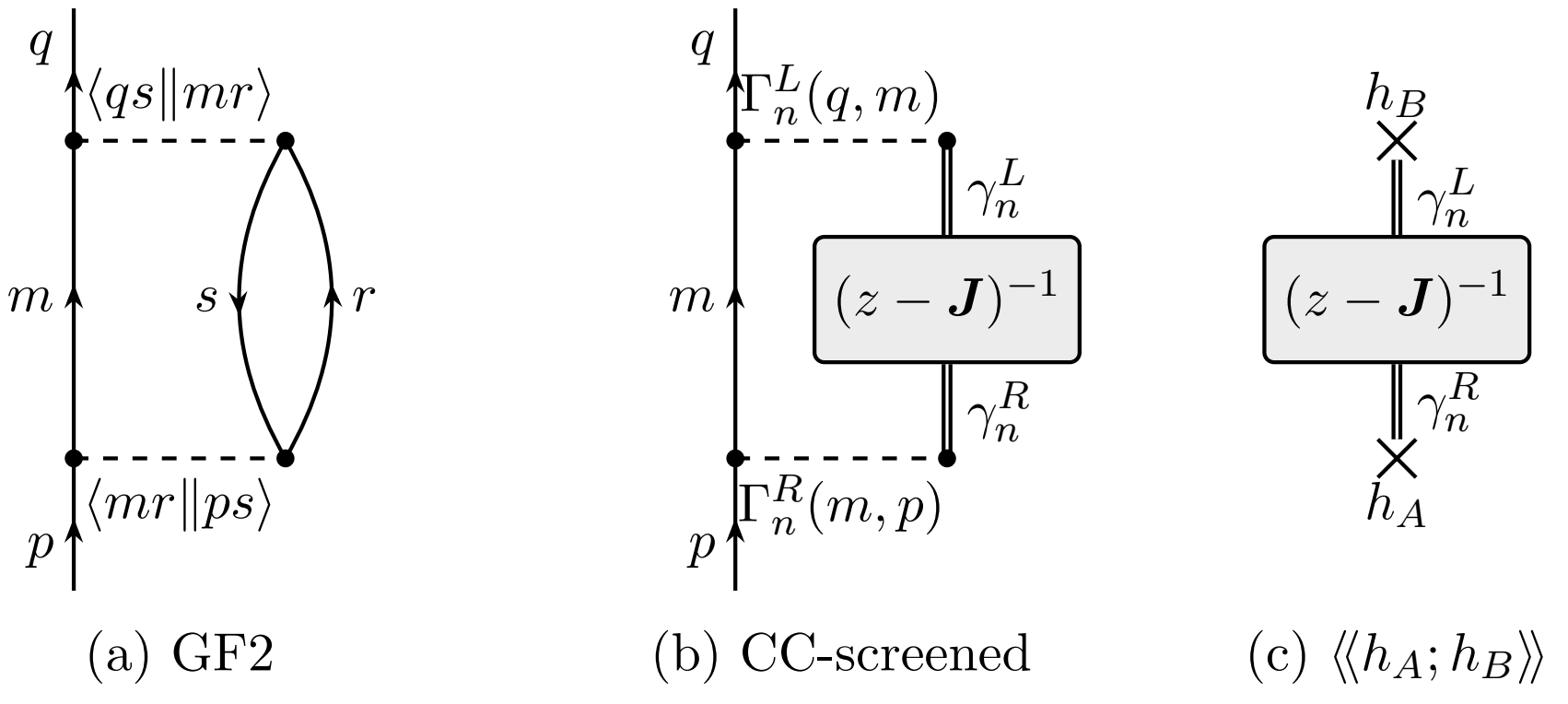
\begin{figure*}
\centering
\begin{tikzpicture}[scale=0.95]
\begin{scope}[shift={(0,0)}]
  \coordinate (b1) at (0,0); \coordinate (b2) at (1.6,0);
  \coordinate (t1) at (0,2.4); \coordinate (t2) at (1.6,2.4);
  \draw[interaction] (b1) -- (b2); \draw[interaction] (t1) -- (t2);
  \draw[fermion] (0,-1.0) -- (b1) node[pos=0.3,left] {$p$};
  \draw[fermion] (b1) -- (t1) node[midway,left] {$m$};
  \draw[fermion] (t1) -- (0,3.4) node[pos=0.7,left] {$q$};
  \draw[fermion] (b2) to[out=60,in=-60] node[right] {$r$} (t2);
  \draw[fermion] (t2) to[out=-120,in=120] node[left] {$s$} (b2);
  \foreach \v in {b1,b2,t1,t2} \node[vertexdot] at (\v) {};
  \node at (0.8,-0.35) {\small $\langle mr\Vert ps\rangle$};
  \node at (0.8,2.75) {\small $\langle qs\Vert mr\rangle$};
  \node at (0.8,-1.5) {(a) GF2};
\end{scope}
\begin{scope}[shift={(5.0,0)}]
  \coordinate (b1) at (0,0); \coordinate (b2) at (1.6,0);
  \coordinate (t1) at (0,2.4); \coordinate (t2) at (1.6,2.4);
  \draw[interaction] (b1) -- (b2); \draw[interaction] (t1) -- (t2);
  \draw[fermion] (0,-1.0) -- (b1) node[pos=0.3,left] {$p$};
  \draw[fermion] (b1) -- (t1) node[midway,left] {$m$};
  \draw[fermion] (t1) -- (0,3.4) node[pos=0.7,left] {$q$};
  \node[respbox] (box) at (1.6,1.2) {\small $(z-\bm{J})^{-1}$};
  \draw[thick,double] (b2) -- (box.south) node[midway,right] {\small $\gamma^R_n$};
  \draw[thick,double] (box.north) -- (t2) node[midway,right] {\small $\gamma^L_n$};
  \foreach \v in {b1,b2,t1,t2} \node[vertexdot] at (\v) {};
  \node at (0.8,-0.35) {\small $\Gamma^R_n(m,p)$};
  \node at (0.8,2.75) {\small $\Gamma^L_n(q,m)$};
  \node at (0.8,-1.5) {(b) CC-screened};
\end{scope}
\begin{scope}[shift={(10.0,0)}]
  \node[respbox] (box) at (0,1.2) {\small $(z-\bm{J})^{-1}$};
  \node[draw,thick,cross out,minimum size=7pt,inner sep=0pt] (hA) at (0,0) {};
  \node[draw,thick,cross out,minimum size=7pt,inner sep=0pt] (hB) at (0,2.4) {};
  \draw[thick,double] (hA) -- (box.south) node[midway,right] {\small $\gamma^R_n$};
  \draw[thick,double] (box.north) -- (hB) node[midway,right] {\small $\gamma^L_n$};
  \node at (0,-0.35) {\small $h_A$}; \node at (0,2.75) {\small $h_B$};
  \node at (0,-1.5) {(c) $\langle\!\langle h_A;h_B\rangle\!\rangle$};
\end{scope}
\end{tikzpicture}
\caption{Antisymmetrized Goldstone diagrams. (a) Second-order self-energy (GF2), drawn for
one time ordering of the intermediate line $m$. (b) The same with the particle--hole pair
replaced by the EOM-CC response: the lower antisymmetrized integral, contracted with the
right transition density, creates the state $n$; it propagates with the EOM-CCSD Jacobian $\bm{J}$; the upper
vertex removes it through the left transition density. (c) The linear-response propagator
of two one-body operators, as used for NMR spin--spin coupling constants; (b) is (c) with
$h_A\to h^{(mp)}$, $h_B\to h^{(qm)}$ [Eq.~\eqref{eq:hpm}].}
\label{fig:diagrams}
\end{figure*}

Figure~\ref{fig:diagrams} shows the corresponding antisymmetrized Goldstone
diagrams.\cite{shavitt2009} In the second-order diagram (a), the dashed lines are
antisymmetrized integrals; the lower one, $\braket{mr||ps}$, scatters the incoming electron
$p$ into $m$ and creates the particle--hole pair $(r,s)$, which is annihilated by the upper
interaction. In (b) the bare pair is replaced by the EOM-CC propagator: the lower vertex
creates the state $R_n$ through $\gamma^R_n$, the state propagates with $\Hbar$, and the
upper vertex removes it through $\gamma^L_n$. Diagram (b) is diagram (c), the
spin--spin-coupling propagator, with the nuclear operators replaced by $h^{(mp)}$ and
$h^{(qm)}$ and the outer line $m$ added (drawn here for the hole ordering). Only one diagram per time ordering of $m$ is
needed because the vertices are antisymmetrized; the corresponding Goldstone expansion
with direct integrals would need the direct and exchange variants of each.

\subsection{Choices of external vertices}

The choice of the external vertices $P$ and $Q$ and of the prefactor $\kappa$ in
Eq.~\eqref{eq:sigma} defines the approximation. We consider three:
\begin{align}
\text{(a)}\quad & P=V,\; Q=V,\; \kappa=1 &&\text{(direct)},\label{eq:var_a}\\
\text{(b)}\quad & P=V,\; Q=\vtx{},\; \kappa=1 &&\text{(mixed direct--antisymmetrized)},\label{eq:var_b}\\
\text{(c)}\quad & P=\vtx{},\; Q=\vtx{},\; \kappa=\tfrac12 &&\text{(antisymmetrized)}.\label{eq:var_c}
\end{align}
We refer to them by these algebraic names. Names from many-body perturbation theory
($GW$, SOSEX, FLEX) are used below only where a reduction has been shown.

In (a) both interactions are direct. The residues of the self-energy are then products of
direct right and left moments, which are the residues of the correlation part of the screened interaction
$W_c=v\chi v$ with $\chi$ the density response of Eq.~\eqref{eq:chiXY}. Variant (a) is
therefore $GW$ with the polarizability of whatever states are supplied; with EOM-CCSD
states this is a $GW$-form self-energy with an EOM-CCSD polarizability, not a
Bethe--Salpeter propagator built from a specified irreducible kernel. Only singlet states
contribute.

In (b) the vertex at the incoming line is direct and the one at the outgoing line
antisymmetrized. The additional term, $-VX$, has the screened (direct) interaction on one side and the bare exchange
interaction on the other. This has the structure of the second-order screened-exchange
(SOSEX) term of Ren \textit{et al.};\cite{ren2015} for direct-RPA states it is exactly
SOSEX (next subsection). With EOM-CCSD states it is the same construction with correlated
states, and we use the algebraic name.

In (c) both interactions are antisymmetrized and all states enter, singlets and triplets
including the spin-flip ($M_S=\pm1$) components. The factor $\tfrac12$ is fixed by the
bare-state limit below, where it gives the correct GF2 symmetry factor: it corrects the
double counting of the two equivalent orientations of the antisymmetrized ring at second
order. At third order the same factor halves every diagram that (c) contains
(Sec.~\ref{sec:pt}).
Structurally, (c) keeps only electron--hole intermediate states, with both vertices
antisymmetrized, and has no particle--particle channel; it resembles the electron--hole
part of a fluctuation-exchange (FLEX) approximation,\cite{bickers1989,marie2025} but with
EOM-CC states rather than a propagator generated from a bare kernel, so we do not
identify the two. Variant (c) is therefore not obtained from (b) simply by adding exchange at the
remaining vertex: the prefactor changes from 1 to $\tfrac12$ and the triplet states enter, so (c)
is a different approximation rather than a systematic extension of (b).

\subsection{Reductions}

\emph{Bare states.} If the states are single orbital-difference excitations
$\ket{\Phi_j^b}$ of the reference, with $\omega_n=\epsilon_b-\epsilon_j$, the transition
densities are $\gamma^R_n(rs)=\delta_{rj}\delta_{sb}$ and $\gamma^L_n(rs)=\delta_{rb}\delta_{sj}$,
and each vertex becomes a single integral. In the hole ordering ($m=i$)
$\vtx{R}_n(i,p)=\braket{ij||pb}$ and $\vtx{L}_n(q,i)=\braket{qb||ij}$; in the particle ordering
($m=a$) $\vtx{R}_n(q,a)=\braket{qj||ab}$ and $\vtx{L}_n(a,p)=\braket{ab||pj}$. The sum over $n$ and
$m$ then runs over all two-hole--one-particle and two-particle--one-hole configurations, and
variant (c) becomes
\begin{equation}
\Sigma^{(2)}_{pq}(\omega)=\frac12\sum_{jab}\frac{\braket{qj||ab}\braket{ab||pj}}{\omega+\epsilon_j-\epsilon_a-\epsilon_b}
+\frac12\sum_{ijb}\frac{\braket{ij||pb}\braket{qb||ij}}{\omega+\epsilon_b-\epsilon_i-\epsilon_j},
\label{eq:gf2}
\end{equation}
the GF2 self-energy, with both terms in the standard index order; the factor $\tfrac12$ appears
because each unordered pair of virtual (or occupied) orbitals is reached from two (state, $m$)
combinations. For variant (b) the $p$ vertex is direct, $V^R_n(i,p)=\braket{ij|pb}$ and
$V^L_n(a,p)=\braket{ab|pj}$, and since
$\sum_{ab}\braket{ab|pj}\braket{qj||ab}=\tfrac12\sum_{ab}\braket{ab||pj}\braket{qj||ab}$ (and
likewise for the occupied pair), variant (b) gives Eq.~\eqref{eq:gf2} as well, while variant (a)
gives only its direct (ring) part.

\emph{Direct-RPA states.} If the states are the singlet solutions of the direct RPA
($A=D+2K$, $B=2K$ in the spin-adapted closed-shell form, $D_{ia,jb}=\delta_{ij}\delta_{ab}(\epsilon_a-\epsilon_i)$,
$K_{ia,jb}=(ia|jb)$), variant (a) is $G_0W_0$@HF exactly. These states are the
EOM-CC states of ring-CCD,\cite{berkelbach2018,rishi2020} whose ground state is the
RPA correlation energy.\cite{scuseria2008} We obtain them by solving
$D^{1/2}(D+4K)D^{1/2}Z=\omega^2 Z$.

For variant (b) the exchange vertex needs the forward and backward amplitudes separately:
the right transition density of state $n$ has $\sqrt2X_{ia,n}$ in its excitation and
$\sqrt2Y_{ia,n}$ in its de-excitation block (spin-summed), and the left density the reverse.
The exchange term of (b), $-V^R_nX^L_n$, then has one bare exchange interaction and the RPA
response on the other side. With $W_p=v\chi v$ and $\chi=\chi_0+\chi v\chi_0$, the product
$W_p\chi_0$ that appears in SOSEX between the screened and the bare exchange interaction equals
$v(\chi-\chi_0)$ in the pair space (equivalently, $v\chi_0W_p=W_p\chi_0v=v(\chi-\chi_0)v$). The
second-order exchange (SOX) term is $\int G\,v\chi_0v_x$ and the SOSEX$(W_p,v)$ term is
$\int G\,W_p\chi_0v_x=\int G\,v(\chi-\chi_0)v_x$, so their sum is $\int G\,v\chi v_x$, which is
the Lehmann form of $-V^R_nX^L_n$. Variant (b) with direct-RPA states is therefore
$G_0W_0$+SOX+SOSEX$(W_p,v)$, i.e.\ $G_0W_0$+SOSEX$(W,v)$ with $W=v+W_p$ in the convention
of Ren \textit{et al.}\cite{ren2015} Variant (c) was not
considered with the singlet-only direct-RPA response used here.

\emph{EOM-CCSD states.} With all EOM-CCSD states, variant (a) is $GW$ with an EOM-CCSD
polarizability, the quantity studied by Lewis and Berkelbach;\cite{lewis2019} (b) makes
one of the two self-energy vertices antisymmetrized; and (c) makes both antisymmetrized,
with CC-correlated electron--hole states. Relative to $G_0W_0$, (a) changes only the
screening and (b) in addition introduces exchange at one vertex of the self-energy; (c), with both
vertices antisymmetrized, the prefactor $\tfrac12$ and the triplet states, is a separate
approximation.
Table~\ref{tab:reductions} summarizes the reductions.

\begin{table}
\caption{Reductions of Eq.~\eqref{eq:sigma} for different states and choices of external vertices.}
\label{tab:reductions}
\begin{ruledtabular}
\begin{tabular}{lccc}
States & (a) direct & (b) mixed & (c) antisymmetrized \\
\hline
bare $\ket{\Phi_j^b}$ & GF2 ring part & GF2 & GF2 \\
direct RPA & $G_0W_0$@HF & $G_0W_0$+SOSEX & --- \\
EOM-CCSD & $GW$, EOM-CCSD $\chi$ & mixed, EOM-CCSD states & antisym., EOM-CCSD states \\
\end{tabular}
\end{ruledtabular}
\end{table}

\subsection{Perturbative structure}\label{sec:pt}

\begin{figure*}
\centering
\begin{tikzpicture}[scale=0.78,every node/.style={scale=0.85}]
\begin{scope}[shift={(0,0)}]
  \coordinate (b1) at (0,0); \coordinate (t1) at (0,3);
  \coordinate (b2) at (1.6,0); \coordinate (t2) at (1.6,3);
  \coordinate (L) at (1.15,1.5); \coordinate (R) at (2.05,1.5);
  \draw[fermion] (0,-0.9) -- (b1) node[pos=0.3,left] {$p$};
  \draw[fermion] (b1) -- (t1) node[midway,left] {$a$};
  \draw[fermion] (t1) -- (0,3.9) node[pos=0.7,left] {$q$};
  \draw[interaction] (b1) -- (b2); \draw[interaction] (t1) -- (t2); \draw[interaction] (L) -- (R);
  \draw[fermion] (b2) to[out=20,in=-90] node[right] {$c$} (R);
  \draw[fermion] (R) to[out=90,in=-20] node[right] {$b$} (t2);
  \draw[fermion] (t2) to[out=-160,in=90] node[left] {$j$} (L);
  \draw[fermion] (L) to[out=-90,in=160] node[left] {$k$} (b2);
  \foreach \v in {b1,b2,t1,t2,L,R} \node[vertexdot] at (\v) {};
  \node at (1.0,-1.5) {(a) ph in $\bm{J}^{(1)}$};
\end{scope}
\begin{scope}[shift={(3.9,0)}]
  \coordinate (b1) at (0,1.0); \coordinate (t1) at (0,3);
  \coordinate (T1) at (0.9,-0.4); \coordinate (T2) at (2.1,-0.4); \coordinate (t2) at (2.1,3);
  \draw[fermion] (0,-0.9) -- (b1) node[pos=0.2,left] {$p$};
  \draw[fermion] (b1) -- (t1) node[midway,left] {$a$};
  \draw[fermion] (t1) -- (0,3.9) node[pos=0.7,left] {$q$};
  \draw[interaction] (T1) -- (T2); \draw[interaction] (t1) -- (t2);
  \draw[fermion] (T1) to[out=100,in=-10] node[right,pos=0.55] {$c$} (b1);
  \draw[fermion] (b1) to[out=-70,in=170] node[left,pos=0.55] {$k$} (T1);
  \draw[fermion] (T2) to[out=60,in=-60] node[right] {$b$} (t2);
  \draw[fermion] (t2) to[out=-120,in=120] node[left] {$j$} (T2);
  \foreach \v in {b1,t1,t2,T1,T2} \node[vertexdot] at (\v) {};
  \node at (1.0,-1.5) {(b) ph, $T_2$ ($\Lambda_2$)};
\end{scope}
\begin{scope}[shift={(7.8,0)}]
  \coordinate (b1) at (0,0); \coordinate (t1) at (0,3);
  \coordinate (b2) at (1.6,0); \coordinate (t2) at (1.6,3);
  \coordinate (E) at (0,1.5); \coordinate (R) at (2.05,1.5);
  \draw[fermion] (0,-0.9) -- (b1) node[pos=0.3,left] {$p$};
  \draw[fermion] (b1) -- (E) node[midway,left] {$a$};
  \draw[fermion] (E) -- (t1) node[midway,left] {$c$};
  \draw[fermion] (t1) -- (0,3.9) node[pos=0.7,left] {$q$};
  \draw[interaction] (b1) -- (b2); \draw[interaction] (t1) -- (t2); \draw[interaction] (E) -- (R);
  \draw[fermion] (b2) to[out=20,in=-90] node[right] {$b$} (R);
  \draw[fermion] (R) to[out=90,in=-20] node[right] {$d$} (t2);
  \draw[fermion] (t2) to[out=-150,in=150] node[left] {$j$} (b2);
  \foreach \v in {b1,b2,t1,t2,E,R} \node[vertexdot] at (\v) {};
  \node at (1.0,-1.5) {(c) pp ladder$^{\dagger}$};
\end{scope}
\begin{scope}[shift={(11.9,0)}]
  \coordinate (V) at (1.0,1.0); \coordinate (t1) at (0,3); \coordinate (t2) at (2.0,3);
  \coordinate (T1) at (0,-0.4); \coordinate (T2) at (2.0,-0.4);
  \draw[fermion] (1.0,-0.9) -- (V) node[pos=0.2,right] {$p$};
  \draw[fermion] (V) to[out=-160,in=30] node[above,pos=0.45] {$k$} (T1);
  \draw[fermion] (V) to[out=-20,in=150] node[above,pos=0.45] {$l$} (T2);
  \draw[interaction] (T1) -- (T2); \draw[interaction] (t1) -- (t2);
  \draw[fermion] (T1) -- (t1) node[midway,left] {$a$};
  \draw[fermion] (T2) -- (t2) node[midway,right] {$b$};
  \draw[fermion] (t2) to[out=-150,in=80] node[left,pos=0.5] {$j$} (V);
  \draw[fermion] (t1) -- (0,3.9) node[pos=0.7,left] {$q$};
  \foreach \v in {V,t1,t2,T1,T2} \node[vertexdot] at (\v) {};
  \node at (1.0,-1.5) {(d) hh-type, $T_2$$^{\dagger}$};
\end{scope}
\begin{scope}[shift={(15.6,0)}]
  \coordinate (V) at (0,1.5); \coordinate (S) at (1.0,1.5);
  \coordinate (T1) at (1.0,0.3); \coordinate (T2) at (1.0,2.7);
  \draw[fermion] (0,-0.9) -- (V) node[pos=0.3,left] {$p$};
  \draw[fermion] (V) -- (0,3.9) node[pos=0.7,left] {$q$};
  \draw[interaction] (V) -- (S);
  \draw[thick] (S) circle[radius=0pt];
  \draw[thick,fill=gray!15] (1.45,1.5) ellipse (0.45 and 0.9);
  \node at (1.45,1.5) {\small $\rho^{(2)}$};
  \node[vertexdot] at (V) {};
  \node at (0.8,-1.5) {(e) static$^{\dagger}$};
\end{scope}
\begin{scope}[shift={(0.6,-6.4)}]
  \begin{scope}[shift={(0,0)}]
    \coordinate (b1) at (0,0); \coordinate (b2) at (1.6,0); \coordinate (t1) at (0,3); \coordinate (t2) at (1.6,3);
    \coordinate (L) at (1.15,1.5); \coordinate (R) at (2.05,1.5);
    \draw[interaction] (b1) -- (b2); \draw[interaction] (t1) -- (t2); \draw[interaction] (L) -- (R);
    \draw[fermion] (0,-0.9) -- (b1) node[pos=0.3,left] {$p$};
    \draw[fermion] (b1) -- (t1) node[midway,left] {$a$};
    \draw[fermion] (t1) -- (0,3.9) node[pos=0.7,left] {$q$};
    \draw[fermion] (b2) to[out=20,in=-90] (R); \draw[fermion] (R) to[out=90,in=-20] (t2);
    \draw[fermion] (t2) to[out=-160,in=90] (L); \draw[fermion] (L) to[out=-90,in=160] (b2);
    \foreach \v in {b1,b2,t1,t2,L,R} \node[vertexdot] at (\v) {};
    \node at (1.0,-1.5) {(f) $dd$};
  \end{scope}
  \begin{scope}[shift={(4.2,0)}]
    \coordinate (b1) at (0,0); \coordinate (b2) at (1.6,0); \coordinate (t1) at (0,3); \coordinate (t2) at (1.6,3);
    \coordinate (L) at (1.15,1.5); \coordinate (R) at (2.05,1.5);
    \draw[interaction] (b1) -- (b2); \draw[interaction] (t1) -- (t2); \draw[interaction] (L) -- (R);
    \draw[fermion] (0,-0.9) -- (b1) node[pos=0.3,left] {$p$};
    \draw[fermion] (b1) -- (t1) node[midway,left] {$a$};
    \draw[fermion] (t1) to[out=-30,in=110] (L);
    \draw[fermion] (R) to[out=90,in=-20] (t2);
    \draw[fermion] (t2) -- (1.6,3.9) node[pos=0.7,right] {$q$};
    \draw[fermion] (b2) to[out=20,in=-90] (R); \draw[fermion] (L) to[out=-90,in=160] (b2);
    \foreach \v in {b1,b2,t1,t2,L,R} \node[vertexdot] at (\v) {};
    \node at (1.0,-1.5) {(g) $dx$};
  \end{scope}
  \begin{scope}[shift={(8.4,0)}]
    \coordinate (b1) at (0,0); \coordinate (b2) at (1.6,0); \coordinate (t1) at (0,3); \coordinate (t2) at (1.6,3);
    \coordinate (L) at (1.15,1.5); \coordinate (R) at (2.05,1.5);
    \draw[interaction] (b1) -- (b2); \draw[interaction] (t1) -- (t2); \draw[interaction] (L) -- (R);
    \draw[fermion] (0,-0.9) -- (b1) node[pos=0.3,left] {$p$};
    \draw[fermion] (b1) to[out=40,in=-100] (R);
    \draw[fermion] (b2) to[out=150,in=-60] node[pos=0.75,left] {$a$} (t1);
    \draw[fermion] (t1) -- (0,3.9) node[pos=0.7,left] {$q$};
    \draw[fermion] (R) to[out=90,in=-20] (t2); \draw[fermion] (t2) to[out=-160,in=90] (L);
    \draw[fermion] (L) to[out=-90,in=120] (b2);
    \foreach \v in {b1,b2,t1,t2,L,R} \node[vertexdot] at (\v) {};
    \node at (1.0,-1.5) {(h) $xd$};
  \end{scope}
  \begin{scope}[shift={(12.6,0)}]
    \coordinate (b1) at (0,0); \coordinate (b2) at (1.6,0); \coordinate (t1) at (0,3); \coordinate (t2) at (1.6,3);
    \coordinate (L) at (1.15,1.5); \coordinate (R) at (2.05,1.5);
    \draw[interaction] (b1) -- (b2); \draw[interaction] (t1) -- (t2); \draw[interaction] (L) -- (R);
    \draw[fermion] (0,-0.9) -- (b1) node[pos=0.3,left] {$p$};
    \draw[fermion] (b1) to[out=40,in=-100] (R);
    \draw[fermion] (R) to[out=90,in=-20] (t2);
    \draw[fermion] (t2) -- (1.6,3.9) node[pos=0.7,right] {$q$};
    \draw[fermion] (b2) to[out=150,in=-60] node[pos=0.75,left] {$a$} (t1);
    \draw[fermion] (t1) to[out=-30,in=110] (L);
    \draw[fermion] (L) to[out=-90,in=120] (b2);
    \foreach \v in {b1,b2,t1,t2,L,R} \node[vertexdot] at (\v) {};
    \node at (1.0,-1.5) {(i) $xx$};
  \end{scope}
\end{scope}
\end{tikzpicture}
\caption{Third-order self-energy diagrams in the particle ordering ($m=a$); the hole ordering
follows by reversing the lines of the intermediate configuration. (a)--(e): antisymmetrized
(Hugenholtz-type) diagrams of the exact third-order self-energy: (a) one particle--hole
interaction $\braket{bk||jc}$ inserted into the bubble, from $\bm{J}^{(1)}$; (b) the bubble opened
into a ground-state double excitation, from $T_2$ (and, at the other end, $\Lambda_2$); (c) the
particle--particle ladder between the intermediate line and the bubble; (d) the coupling
correction with a hole--hole pair of $T_2$; (e) the static term, with $\rho^{(2)}$ the
second-order density. $^{\dagger}$Not contained in any of the three variants. (f)--(i): the
four Goldstone forms of (a) at the outer vertices (direct $d$ or exchanged $x$ attachment of the
external line; first letter lower vertex, here the $p$ vertex): (f) direct at both, (g) direct at
$p$ and exchanged at $q$, (h) exchanged at $p$ and direct at $q$, (i) exchanged at both; (b) splits
in the same way. Table~\ref{tab:pt3} gives the weights.}
\label{fig:pt3}
\end{figure*}
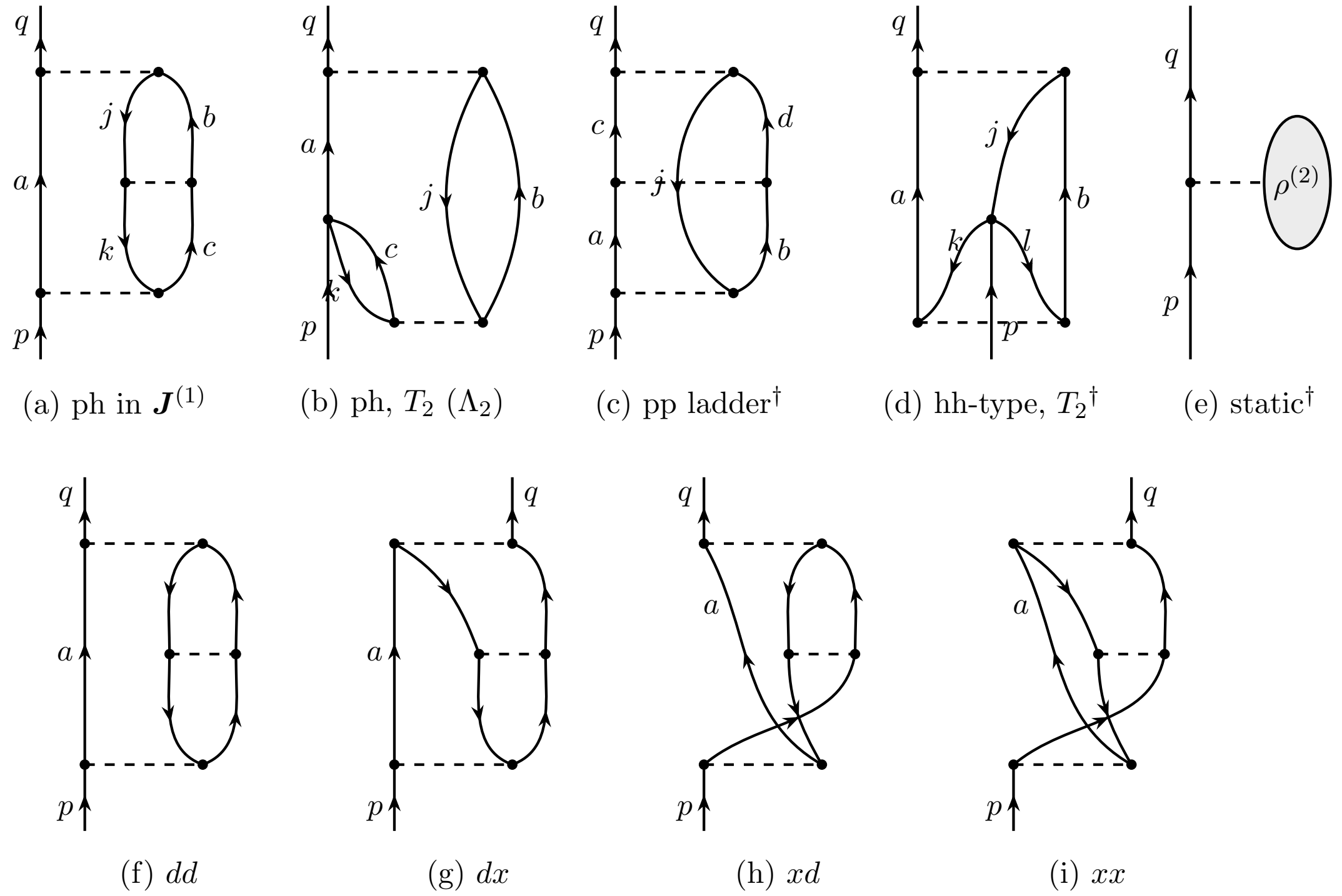

To identify the diagrams in Eq.~\eqref{eq:sigma} beyond these limits we keep the canonical
Hartree--Fock orbitals fixed and scale the fluctuation potential,
\begin{equation}
H(\lambda)=F+\lambda W,\qquad W=H-F,
\label{eq:Hlambda}
\end{equation}
where $F=E_\mathrm{HF}+\sum_p\epsilon_p\{a^\dagger_pa_p\}$ is the Fock operator of the frozen
Hartree--Fock reference and $W=V_N$ is the two-electron interaction normal-ordered with respect to
it, so that the one-body mean field of the scaled interaction is subtracted and the reference
energy enters only as a constant (Appendix~\ref{app:resp}). The CC
amplitudes then start as $T_2=\lambda T_2^{(1)}+\dots$, $T_1=O(\lambda^2)$, and
$\Lambda_2^{(1)}=T_2^{(1)\dagger}$. Each outer vertex carries one factor $\lambda$, so the
self-energy starts at second order, and its $n$th-order term is fixed by the response to order
$n-2$. We expand the response as a resolvent rather than state by state. For one-body operators
$X$ and $Y$ the forward part of Eq.~\eqref{eq:chiXY} is
\begin{equation}
\langle\!\langle X;Y\rangle\!\rangle^{+}_z=\bm{l}_X\,(z-\bm{J})^{-1}\,\bm{r}_Y,\qquad
l_X(\mu)=\bra{0}(1+\Lambda)\bar X\tau_\mu\ket{0}-\langle\bar X\rangle\,\Lambda_\mu,\quad r_Y(\mu)=\bra{\mu}\bar Y\ket{0},
\label{eq:resolvent}
\end{equation}
where $\tau_\mu$ runs over single and double excitations,
$J_{\mu\nu}=\bra{\mu}\Hbar-E_\mathrm{CC}\ket{\nu}$ is the EOM-CCSD Jacobian,
$\langle\bar X\rangle=\bra{0}(1+\Lambda)\bar X\ket{0}$ and $\Lambda_\mu=\bra{0}\Lambda\tau_\mu\ket{0}$.
Inserting $\sum_nR_nL_n$ for the resolvent gives back the moments of Eqs.~\eqref{eq:gammaR}
and~\eqref{eq:gammaL}; the last term of $l_X$ is the reference component of $R_n$ in
Eq.~\eqref{eq:gammaR}, $r_{0,n}\langle\bar X\rangle$ with $r_{0,n}=-\bra{0}\Lambda R_n\ket{0}$.
It cancels the disconnected part $\bra{0}\bar X\ket{0}\Lambda_\mu$ of the first term, leaving
$-\bra{0}\Lambda\bar X\ket{0}\Lambda_\mu$, with $\bra{0}\Lambda\bar X\ket{0}=O(\lambda^2)$ beyond the vertex factor; this term first enters the
response at fourth order (the self-energy at sixth order); it therefore does not affect the
second- and third-order analysis below. With $\bm{J}=\bm{J}^{(0)}+\lambda\bm{J}^{(1)}+\dots$ and
$\bm{R}_0(z)=(z-\bm{J}^{(0)})^{-1}$, where $\bm{J}^{(0)}$ is diagonal in orbital-energy differences,
\begin{equation}
\langle\!\langle X;Y\rangle\!\rangle^{+(1)}=
\bm{l}^{(0)}_X\bm{R}_0\bm{J}^{(1)}\bm{R}_0\,\bm{r}^{(0)}_Y
+\bm{l}^{(1)}_X\bm{R}_0\,\bm{r}^{(0)}_Y+\bm{l}^{(0)}_X\bm{R}_0\,\bm{r}^{(1)}_Y .
\label{eq:chi1}
\end{equation}
The zeroth-order moments of a double excitation vanish, so to this order only the singles
block enters: $J^{(1)}_{jb,kc}=\braket{bk||jc}$ is the particle--hole interaction,
$r^{(1)}_Y(jb)=\sum_{kc}Y_{kc}t^{bc}_{jk}$ is the de-excitation part of $Y$ acting on the
first-order ground state, and $l^{(1)}_X(jb)=\sum_{kc}X_{ck}\lambda^{jk}_{bc}$ is the excitation
part of $X$ closed by $\Lambda_2$ (Appendix~\ref{app:resp}). The zeroth-order response gives the
bare-state reduction above; Eq.~\eqref{eq:chi1}, attached to the outer vertices, gives all
third-order terms of each variant. In the particle ordering, for example, the first term gives
\begin{equation}
\Sigma^{(3)}_{J,pq}=\kappa\sum_a\sum_{jb,kc}
\frac{X_{jb}\,\braket{bk||jc}\,Y_{ck}}{(\omega+\epsilon_j-\epsilon_a-\epsilon_b)(\omega+\epsilon_k-\epsilon_a-\epsilon_c)},
\label{eq:sigJ}
\end{equation}
with $X=h^{(qa)}$ and $Y=h^{(ap)}$ (antisymmetrized or direct, according to the variant); the
coefficient of each exact diagram is derived from this expression in Appendix~\ref{app:sigma3}.
Diagrammatically, the first term inserts one particle--hole interaction into the bubble, and the
other two open the bubble at one end into a ground-state double excitation (Fig.~\ref{fig:pt3}).

The exact third-order self-energy, in the decomposition of the algebraic diagrammatic
construction,\cite{schirmer1983} has three parts (Appendix~\ref{app:sigma3}): the first-order
coupling among the 2p1h (or 2h1p) configurations, which contains the particle--hole
interactions and the particle--particle (hole--hole) ladders between the intermediate line and
the particle (hole) of the pair; the second-order coupling corrections, which contain one
ground-state amplitude $t_2$ and are again of particle--hole or of pp/hh-ladder type; and a
static term, fixed by the second-order density. All three follow from the third-order expansion of
the Dyson ADC(3) self-energy (Appendix~\ref{app:sigma3}). Each particle--hole diagram can be split at its two outer
vertices into the four Goldstone diagrams with direct ($d$) or exchanged ($x$) attachment of the
external line [Fig.~\ref{fig:pt3}(f)--(i)]. In $dd$ the interaction acts inside the bubble, so
$dd$ collects the third-order corrections to the polarizability (the direct ring and the
particle--hole exchange terms of the kernel); in $xx$ it connects the external line with the bubble, which is a vertex correction of
the self-energy; $dx$ and $xd$ are mixed.

Table~\ref{tab:pt3} gives the weight with which each variant contains each exact third-order
diagram. Variant (a) contains all polarizability diagrams with their exact weight and nothing
else: with EOM-CCSD states it is, to third order, $GW$ with the correlated polarizability and no
vertex correction. Variant (b) contains, with full weight, every particle--hole diagram that is direct at the vertex
of the incoming line $p$, i.e.\ in addition the mixed diagrams exchanged at the $q$ vertex, and
none exchanged at the $p$ vertex; for $\Sigma_{pp}$ the two kinds of mixed diagram are equal, so
(b) has half of the total mixed contribution. Variant (c)
contains every particle--hole diagram, but each with half its weight: the factor $\tfrac12$ is
correct at second order, where $dd$ and $xx$ are the same diagram, and is applied at third order
to diagrams that are no longer equivalent. At third order, none of the variants contains the explicit
particle--particle and hole--hole ladders involving the intermediate line $m$, the corresponding
pp/hh-type coupling corrections, or the static term. The intermediate line $m$ is a bare $G^0$
that does not interact with the electron--hole pair, and $\Sigma$ falls off as $\omega^{-1}$.
The weights follow algebraically from the contractions of Eq.~\eqref{eq:chi1} with the outer
vertices (Appendix~\ref{app:sigma3}, worked example).

The self-energy of Eq.~\eqref{eq:sigma} is not the self-energy of the coupled-cluster
one-particle Green's function, which is built from ionized and attached
states;\cite{nooijen1992,coveney2025} the two need not agree beyond second order. The
diagrammatic analysis above holds through third order. Beyond it, the CC response contains
further ground-state, Jacobian and transition-moment contributions, and we do not claim
equivalence to any particular diagrammatic vertex approximation.

\begin{table}
\caption{Third-order diagram content of the three variants, as weights relative to the exact
third-order self-energy. Particle--hole (ph) diagrams are classified by the attachment at the two
outer vertices, direct or exchanged (Fig.~\ref{fig:pt3}); a vertex counts as exchanged if any
element attached there (the outer coupling, or the integral of a ground-state correction located
at that vertex) is exchanged. The rows apply alike to the ph interaction from $\bm{J}^{(1)}$, to the
ph coupling corrections from $T_2$ and $\Lambda_2$, to the direct (ring) and exchange
parts of $\braket{bk||jc}$, to both time orderings and to $\Sigma_{pq}$ with $p\neq q$.}
\label{tab:pt3}
\begin{ruledtabular}
\begin{tabular}{lcccc}
Third-order diagrams & exact & (a) & (b) & (c) \\
\hline
ph, direct at $p$ and $q$ ($dd$, polarizability) & 1 & 1 & 1 & $\tfrac12$ \\
ph, direct at $p$, exchanged at $q$ & 1 & 0 & 1 & $\tfrac12$ \\
ph, exchanged at $p$, direct at $q$ & 1 & 0 & 0 & $\tfrac12$ \\
ph, exchanged at $p$ and $q$ ($xx$, self-energy vertex) & 1 & 0 & 0 & $\tfrac12$ \\
pp and hh ladders (with the intermediate line) & 1 & 0 & 0 & 0 \\
pp/hh-type coupling corrections & 1 & 0 & 0 & 0 \\
static & 1 & 0 & 0 & 0 \\
\end{tabular}
\end{ruledtabular}
\end{table}

\subsection{Quasiparticle equation}

With a Hartree--Fock reference the exchange part of the self-energy cancels against the
Fock operator, and the diagonal quasiparticle equation is
\begin{equation}
\omega=\epsilon_p+\Sigma_{pp}(\omega).
\label{eq:qp}
\end{equation}
Because $\Sigma_{pp}(\omega)=\sum_k c_k/(\omega-P_k)$ is rational, its roots can be searched
for systematically. When all residues $c_k$ are real and non-negative (variant (a) with bare or
direct-RPA states), there is exactly one root between neighbouring poles. Since
$\Sigma_{pp}(\omega)=O(\omega^{-1})$, $G_{pp}=[\omega-\epsilon_p-\Sigma_{pp}]^{-1}$ behaves as
$\omega^{-1}$ for large $\omega$, so its residues, the spectral weights
$Z=[1-\partial\Sigma_{pp}/\partial\omega]^{-1}$ at the roots, sum to one. The quasiparticle
is the root of largest weight; the others are satellites. The residues $c_k$ of the self-energy
and the weights $Z$ of the Green's function are different quantities. For EOM-CC states, and
in (b) and (c) in general, the $c_k$ are products of left and right quantities of a
non-Hermitian problem and need not be positive; a negative self-energy residue can, but need
not, lead to a negative weight in $G$. The sum rule still holds, but causality and a
non-negative spectral function are not guaranteed. In this respect the construction differs
from ADC-type and positive-semidefinite vertex-corrected self-energies, which are built to have
the Lehmann form with positive semi-definite residues;\cite{marie2026adc,bruneval2025psd,pavlyukh2026psd}
in Eq.~\eqref{eq:sigma} the sum-over-states form is inherited from $\Hbar$, but the positivity of
the residues is not. Because $\Hbar$ is not Hermitian, a complete diagonalization can also give
complex-conjugate pairs of eigenvalues. We use $\Sigma$ only at real $\omega$, for the
quasiparticle equation; the size of these effects in the systems studied is given in
Sec.~\ref{sec:analytic}.

Eq.~\eqref{eq:qp} uses only the diagonal of $\Sigma$, but the off-diagonal elements tell
whether the variants define a symmetric matrix, which matters when the full
$\Sigma_{pq}$ is needed (e.g.\ for a non-diagonal or self-consistent solution). Variant (a) with
direct-RPA states is symmetric; with EOM-CCSD states (a) and (c) are asymmetric only through the
non-Hermiticity of $\Hbar$. Variant (b) is structurally asymmetric: its incoming vertex is direct
and its outgoing vertex antisymmetrized, so $\Sigma_{pq}$ contains the mixed diagrams of one kind
($dx$, Sec.~\ref{sec:pt}) and $\Sigma_{qp}$ those of the other ($xd$), which are equal only for
$p=q$. The diagonal elements, and so all results reported here, are unaffected. Where the full
matrix is needed, a possible heuristic is the symmetric part $\tfrac12(\Sigma_{pq}+\Sigma_{qp})$
of (b), which restores symmetry at real frequencies but guarantees neither causality nor a
positive spectral weight; averaging the two mixed-vertex orientations would instead define a
different approximation.

\section{Computational details}\label{sec:comp}
All calculations used the ACES~II program system (development version ACES\_PROGRAM),
modules for CCSD, $\Lambda$-CCSD, EOM-CCSD and the self-energy described here.
The reference is UHF, which for the closed-shell systems studied is the RHF determinant; all
electrons are correlated; the basis is cc-pVDZ; geometries are experimental (NIST CCCBDB; coordinates in the Supplementary Material).
The systems are Ne, HF, H$_2$O, NH$_3$, CH$_4$, BF, CO and N$_2$.

The EOM-EE-CCSD similarity-transformed Hamiltonian in the space of single and double
excitations is diagonalized completely in each irreducible representation of the abelian
point group, giving all right and left eigenvectors (2835 states for Ne up to 30\,723 for BF,
CO and N$_2$). For each state the right and left one-particle transition densities
$\gamma^R_n$ and $\gamma^L_n$ are formed for both spins, and the direct and exchange vertices
of Eq.~\eqref{eq:vertices} are built from the full set of molecular-orbital integrals.
Spin-flip ($M_S=\pm1$) states, needed only in variant (c), enter through spin isotropy from
the triplet part of the $M_S=0$ states. The direct-RPA states (one singlet per
occupied--virtual pair) are obtained by diagonalizing $D^{1/2}(D+4K)D^{1/2}$ in the same code.
Only the left and right states of $\Hbar$ enter; no frequency grid, quadrature or analytic
continuation is used for the self-energy.

Because $\Sigma_{pp}$ is rational, its poles are known and Eq.~\eqref{eq:qp} can be searched
interval by interval. As a practical procedure for identifying the principal quasiparticle
solution, for each occupied orbital and each variant we take the root of largest weight $Z$ among (i) the roots bracketed within $\epsilon_p\pm0.5$~$E_h$, each pole
interval being sampled at interior points because negative residues allow pairs of roots
without a sign change between poles, and (ii) the root reached by Newton iteration from
$\epsilon_p$ (step capped at 0.05~$E_h$, converged to $10^{-10}$~$E_h$). Using both guards
against a root missed by either search. The procedure is not an exhaustive enumeration: with negative residues some roots of the rational equation can be
complex, and between two poles neither the number of real roots nor their monotonic ordering is
guaranteed, so we have not established that every real root is found. For the 20 outer-valence
levels the selected root carries $Z=0.77$--0.95 in (a), 0.79--0.96 in (b), 0.65--0.92 in (c) and
0.57--0.97 in $G_0W_0$@HF; for the inner-valence levels the weight is shared among several
roots and the selection is not unambiguous (Sec.~\ref{sec:inner}). For the outer-valence levels the
choice matters in one case: for the HF 2$\sigma$ level in variant (c) the Newton iteration
converges to a satellite with $Z=0.03$. The reference IPs are from a complete diagonalization of IP-EOM-CCSD\cite{stanton1994} in the same
basis without symmetry. The main line of orbital $p$ is the root with the largest one-hole
weight $P_p$; for a degenerate set of orbitals $P$ is summed over the set. Statistics use the
20 levels whose IP-EOM-CCSD main-line weight exceeds 0.8. The remaining six (inner-valence
levels of H$_2$O, NH$_3$, BF, CO and N$_2$, and the 1$\pi$ level of BF) are listed but not
included.

\emph{Implementation checks.} The limiting cases of Sec.~\ref{sec:theory} were confirmed
numerically: with bare states variants (b) and (c) reproduce GF2, and with direct-RPA states
variants (a) and (b) reproduce $G_0W_0$@HF and $G_0W_0$+SOSEX from an independent implementation,
in all cases to machine precision. The third-order expressions of Sec.~\ref{sec:pt} and
Appendix~\ref{app:sigma3} were checked for H$_2$O (STO-3G, 6-31G) and BeH$_2$ (STO-3G) against the
production code run with scaled two-electron integrals and against the exact self-energy from
full configuration interaction. Details are given in the Supplementary Material.

\section{Results}\label{sec:results}
\subsection{Principal ionization potentials}

The main result is in the error statistics of Table~\ref{tab:stats}, arranged by screening
and self-energy vertex in Table~\ref{tab:grid}; Table~\ref{tab:ips} lists the individual IPs
and the errors of the five self-energies. Changing the screening from direct
RPA to EOM-CCSD lowers every IP (by 0.25--1.26~eV), on average: by $0.67$~eV without and $0.45$~eV with the exchange vertex
(change in mean signed error). Adding the exchange vertex to the self-energy raises every IP, on
average by $0.53$~eV with RPA and $0.75$~eV with CCSD screening. The two changes are of similar size and
opposite sign and are close to additive (the cross term is about 0.2~eV). $G_0W_0$@HF itself
(MSE $+0.27$~eV, MAE $0.31$~eV) and the CC-screened form with the exchange vertex, variant (b)
(MSE $+0.35$~eV, MAE $0.39$~eV), are the two combinations in which the errors balance. Each
correction applied alone moves the result past the reference: $GW$ with the CCSD
polarizability, variant (a), underestimates the IPs (MSE $-0.40$~eV), in line with Lewis and
Berkelbach,\cite{lewis2019} and $G_0W_0$+SOSEX@HF overestimates them (MSE $+0.80$~eV).

The pattern is systematic rather than an average effect. Variant (a) lies below IP-EOM-CCSD
for 17 of the 20 levels (the exceptions, CO 5$\sigma$ and N$_2$ 2$\sigma_u$ and 3$\sigma_g$, are
within 0.1~eV), and $G_0W_0$+SOSEX@HF lies above it for all levels except N$_2$ 1$\pi_u$. The
exchange vertex has its largest effect on the deepest valence levels (Ne 2s, HF 2$\sigma$, CH$_4$
2a$_1$), which it raises by 1.0--2.4~eV. Variant (b) and $G_0W_0$@HF agree to within 0.1~eV for 11 of
the 20 levels. Variant (b) is clearly worse for Ne 2s and HF 2$\sigma$ ($+0.79$ and $+0.96$~eV,
against $-0.01$ and $-0.15$~eV) and better for BF 4$\sigma$, CH$_4$ 1t$_2$ and the 1b$_2$ and 1e
levels of H$_2$O and NH$_3$; both are too high for the N$_2$ $\sigma$ levels.

Variant (c), with both vertices antisymmetrized and all singlet, triplet and spin-flip states,
is much worse (MSE $-1.03$~eV, MAE $1.16$~eV, maximum error 3.0~eV), and lies below the
reference for 18 of the 20 levels. Treating the
electron--hole channel with full exchange but without the particle--particle channel and
without coupling between channels does not help, in agreement with the findings of Marie and
Loos for uncoupled channels of the parquet equations.\cite{marie2025}

\begin{table}
\caption{Errors relative to IP-EOM-CCSD for the 20 levels of Table~\ref{tab:ips} with main-line
weight above 0.8 (eV): mean signed (MSE), mean absolute (MAE) and maximum absolute error. The last
three rows add the static self-energy of Eq.~\eqref{eq:sigstat} (Sec.~\ref{sec:static}).}
\label{tab:stats}
\begin{ruledtabular}
\begin{tabular}{llrrr}
States & Self-energy & MSE & MAE & Max \\
\hline
direct RPA & $G_0W_0$ & $+0.27$ & 0.31 & 0.97 \\
direct RPA & $G_0W_0$+SOSEX & $+0.80$ & 0.81 & 2.22 \\
EOM-CCSD & (a) direct & $-0.40$ & 0.42 & 1.14 \\
EOM-CCSD & (b) mixed & $+0.35$ & 0.39 & 1.40 \\
EOM-CCSD & (c) antisymmetrized & $-1.03$ & 1.16 & 3.00 \\
\hline
direct RPA & $G_0W_0$ + $\Sigma^{(\infty)}$ & $+0.10$ & 0.28 & 0.74 \\
EOM-CCSD & (a) + $\Sigma^{(\infty)}$ & $-0.56$ & 0.56 & 1.02 \\
EOM-CCSD & (b) + $\Sigma^{(\infty)}$ & $+0.18$ & 0.41 & 1.06 \\
\end{tabular}
\end{ruledtabular}
\end{table}

\begin{table}
\caption{Mean signed errors (eV) of Table~\ref{tab:stats} arranged by screening and
self-energy vertex. Each step to the right adds the exchange vertex to the self-energy; each
step down replaces the RPA polarizability by the EOM-CCSD one.}
\label{tab:grid}
\begin{ruledtabular}
\begin{tabular}{lrrr}
 & direct vertex & with exchange & change \\
\hline
RPA & $+0.27$ & $+0.80$ & $+0.53$ \\
EOM-CCSD & $-0.40$ & $+0.35$ & $+0.75$ \\
change & $-0.67$ & $-0.45$ & \\
\end{tabular}
\end{ruledtabular}
\end{table}

\begin{table*}
\caption{Principal ionization potentials (eV): Koopmans values ($-\epsilon_p$),
IP-EOM-CCSD, and errors (eV) relative to IP-EOM-CCSD of $G_0W_0$@HF, $G_0W_0$+SOSEX@HF
(variant (b) with direct-RPA states), and variants (a)--(c) with EOM-CCSD states. cc-pVDZ, all
electrons correlated. $^*$IP-EOM-CCSD main-line weight below 0.8; not included in
Table~\ref{tab:stats}.}
\label{tab:ips}
\renewcommand{\arraystretch}{0.8}
\begin{ruledtabular}
\begin{tabular}{llrrrrrrr}
 & & & & \multicolumn{2}{c}{RPA states} & \multicolumn{3}{c}{EOM-CCSD states} \\
\cline{5-6}\cline{7-9}
System & Level & $-\epsilon_p$ & IP-EOM & $G_0W_0$ & +SOSEX & (a) & (b) & (c) \\
\hline
Ne & 2s & 52.21 & 48.34 & $-$0.01 & +1.45 & $-$1.14 & +0.79 & $-$1.46 \\
 & 2p & 22.64 & 20.62 & +0.24 & +0.79 & $-$0.43 & +0.32 & $-$1.12 \\
\hline
HF & 2$\sigma$ & 43.09 & 38.92 & $-$0.15 & +2.22 & $-$0.99 & +0.96 & $-$1.49 \\
 & 3$\sigma$ & 20.33 & 19.32 & +0.18 & +0.49 & $-$0.31 & +0.18 & $-$0.74 \\
 & 1$\pi$ & 17.11 & 15.20 & +0.33 & +0.83 & $-$0.36 & +0.33 & $-$1.46 \\
\hline
H$_2$O & 2a$_1$ & 36.37 & 32.17$^*$ & +1.21 & +2.43 & $-$1.25 & +1.69 & $-$2.27 \\
 & 1b$_2$ & 19.03 & 18.47 & +0.09 & +0.28 & $-$0.35 & $-$0.04 & $-$0.58 \\
 & 3a$_1$ & 15.42 & 14.11 & +0.32 & +0.66 & $-$0.24 & +0.27 & $-$0.99 \\
 & 1b$_1$ & 13.42 & 11.80 & +0.36 & +0.78 & $-$0.28 & +0.30 & $-$1.32 \\
\hline
NH$_3$ & 2a$_1$ & 30.80 & 27.50$^*$ & +0.43 & +1.91 & $-$1.53 & +1.27 & $-$0.35 \\
 & 1e & 16.77 & 16.17 & +0.17 & +0.33 & $-$0.27 & $-$0.01 & $-$0.48 \\
 & 3a$_1$ & 11.40 & 10.24 & +0.35 & +0.65 & $-$0.19 & +0.26 & $-$0.84 \\
\hline
CH$_4$ & 2a$_1$ & 25.67 & 23.23 & +0.45 & +1.42 & $-$0.81 & +0.65 & $-$1.06 \\
 & 1t$_2$ & 14.78 & 14.19 & +0.24 & +0.37 & $-$0.18 & +0.04 & $-$0.28 \\
\hline
BF & 3$\sigma$ & 46.36 & 41.68$^*$ & $-$0.04 & +1.80 & $-$0.64 & +0.85 & $-$2.16 \\
 & 4$\sigma$ & 23.24 & 21.03 & +0.31 & +0.82 & $-$0.87 & $-$0.09 & $-$2.67 \\
 & 1$\pi$ & 20.19 & 17.91$^*$ & +0.50 & +1.08 & $-$0.81 & +0.50 & $-$1.94 \\
 & 5$\sigma$ & 10.94 & 11.04 & +0.05 & +0.14 & $-$0.20 & +0.08 & +0.61 \\
\hline
CO & 3$\sigma$ & 41.44 & 36.25$^*$ & +0.83 & +3.02 & $-$1.03 & +3.17 & $-$0.08 \\
 & 4$\sigma$ & 21.83 & 19.46 & +0.49 & +1.10 & $-$0.54 & +0.58 & $-$3.00 \\
 & 1$\pi$ & 17.26 & 16.74 & $-$0.02 & +0.28 & $-$0.58 & +0.01 & $-$1.22 \\
 & 5$\sigma$ & 14.93 & 13.81 & +0.58 & +0.85 & +0.03 & +0.51 & $-$0.04 \\
\hline
N$_2$ & 2$\sigma_g$ & 40.03 & 38.55$^*$ & $-$2.29 & $-$0.65 & $-$3.29 & $-$2.16 & $-$1.51 \\
 & 2$\sigma_u$ & 21.06 & 18.47 & +0.97 & +1.65 & +0.06 & +1.40 & $-$2.13 \\
 & 3$\sigma_g$ & 17.04 & 15.18 & +0.68 & +1.08 & +0.09 & +0.76 & $-$1.05 \\
 & 1$\pi_u$ & 16.55 & 16.93 & $-$0.20 & $-$0.11 & $-$0.47 & $-$0.31 & +0.64 \\

\end{tabular}
\end{ruledtabular}
\end{table*}

\emph{Excluded levels.}\label{sec:inner} Six levels are excluded from the statistics because
their IP-EOM-CCSD main line carries less than 0.8 of the one-hole weight: the inner-valence
levels (2a$_1$ of H$_2$O and NH$_3$, 3$\sigma$ of BF and CO, 2$\sigma_g$ of N$_2$) and the
outer-valence 1$\pi$ level of BF. For them the errors of all self-energies are large and erratic
(up to 3.3~eV, N$_2$ 2$\sigma_g$ in variant (a)). The diagonal quasiparticle equation then has
several roots of comparable weight, and the root of largest weight need not be the one that
corresponds to the IP-EOM-CCSD main line; a comparison of the full spectral functions would be
needed and is outside the scope of this paper.

Table~\ref{tab:mag} (Appendix~\ref{app:sigma3}) shows how large the third-order parts are for
H$_2$O in STO-3G, at $\omega=\epsilon_p$. For the outer-valence levels the particle--hole
coupling term (0.3--1.5~eV) and the explicit particle--particle and hole--hole ladders ($-0.2$ to
$-1.2$~eV) are of the same size and of opposite sign, and the exact third order ($-0.14$ to
$-0.38$~eV) is the small remainder. Variants (a) and (b), which contain the polarizability
diagrams with full weight, give the sign of the exact third order for 3a$_1$ and 1b$_1$; variant
(c), which contains half of every particle--hole diagram and none of the ladders, has the wrong
sign for both ($+0.22$ and $+0.49$~eV). In the fully antisymmetrized electron--hole form the
missing ladders are therefore not a small omission: they would cancel most of the
particle--hole terms that (c) keeps. This is consistent with the systematic underestimation of
the IPs by (c) in Table~\ref{tab:ips}, although third order alone does not determine the
resummed result. The problem of (c) is, however, not only the missing ladders: as shown in
Sec.~\ref{sec:pt}, the particle--hole diagrams it retains also carry incorrect third-order
weights (each $\tfrac12$), so adding the ladders alone would not recover the exact third-order
self-energy. What matters is the balance between the contributions.

\subsection{Analytic properties of the computed self-energies}\label{sec:analytic}

To measure the departure from positive self-energy residues we computed all residues of
$\Sigma_{pp}$, summed those at coincident poles (degenerate states, and the two orientations of
one electron--hole pair), and define $W_-$ as the fraction of the total absolute residue weight
of $\Sigma_{pp}$ carried by negative self-energy residues; it is not a negative spectral weight
of $G$. For the valence levels of the eight cc-pVDZ systems, variant (a) with bare or direct-RPA
states has no negative residues, as it must. Variant (b) with direct-RPA states, i.e.\
$G_0W_0$+SOSEX@HF, already has $W_-=1.4$--4.5\%: the negative residues come from the mixed
vertex, not from the CC states. With EOM-CCSD states $W_-$ is $4\times10^{-5}$--2.4\% for (a)
(largest for BF and H$_2$O), 2.6--5.5\% for (b), and $5\times10^{-6}$--0.8\% for (c). Leaving
out the states of complex-conjugate pairs changes these values little, so the pairs are not the
source.

Complex-conjugate pairs of EOM-CCSD excitation energies occur in six of the eight systems
(3 to 37 pairs, $|\mathrm{Im}\,\omega_n|\le1.4\times10^{-3}$~$E_h$). Treating each pair as two
real states at $\mathrm{Re}\,\omega_n$ is not exact, so we also evaluated Eq.~\eqref{eq:sigma}
with each pair entering through its exact $2\times2$ block of $\Hbar$ in the space of the two
stored vectors. The occupied quasiparticle energies change by at most 0.4~meV (CO); only virtual
roots close to a pole, with $Z<0.2$, move appreciably.

For the off-diagonal elements, (a) with direct-RPA states is symmetric to machine precision for
all valence occupied pairs of the same symmetry at $\omega=\mu$. With EOM-CCSD states the
asymmetry of (a) and (c) is at most $7\times10^{-4}$ and $4\times10^{-3}$~eV, against
$|\Sigma_{pq}|$ up to 1.2~eV. The structural asymmetry of (b) reaches 0.08~eV with direct-RPA
states and 0.16~eV (N$_2$, $|\Sigma_{pq}|=0.2$~eV) with EOM-CCSD states.

\subsection{Static self-energy}\label{sec:static}

None of the three forms contains the static part of the exact self-energy,
\begin{equation}
\Sigma^{(\infty)}_{pq}=\sum_{rs}\braket{pr||qs}\,(\gamma-\gamma^\mathrm{HF})_{sr},
\label{eq:sigstat}
\end{equation}
the high-frequency limit of the correlation self-energy, i.e.\ the correlation-induced
correction to the static Hartree--Fock potential, given exactly by the correlated one-body
density $\gamma$.\cite{schirmer1983,schirmer2018} Because $\gamma-\gamma^\mathrm{HF}$ starts at
second order in the fluctuation potential, $\Sigma^{(\infty)}$ starts at third order, and its
leading term is precisely the static contribution that the perturbative analysis of
Sec.~\ref{sec:pt} identifies as missing from all three forms [Eq.~\eqref{eq:sigS}]. Adding
Eq.~\eqref{eq:sigstat} therefore supplies that class of diagrams, and beyond third order its
higher-order density contributions. Corrected density matrices of this kind were used to improve
$GW$ IPs by Bruneval,\cite{bruneval2019prb,bruneval2019jctc} and the same term appears in the
connection between $GW$ and extended CC, where it was found to improve IPs, in particular
together with vertex corrections.\cite{tolle2026} It enters the quasiparticle equation as
$\omega=\epsilon_p+\Sigma^{(\infty)}_{pp}+\Sigma_{pp}(\omega)$ and leaves the poles of
$\Sigma(\omega)$ unchanged. We evaluate it with the CCSD one-body density at fixed reference
(orbital-unrelaxed), consistent with the response that defines the states. An independent
evaluation with the second-order (MP2) density gives the same signs except for two small CO
values (4$\sigma$ and 1$\pi$), and mostly larger magnitudes, so the higher-order part of the
density is not negligible. For the
outer-valence levels $\Sigma^{(\infty)}_{pp}$ ranges from $-0.27$ to $+0.83$~eV and has no
fixed sign: it is negative, and raises the IPs, for Ne, HF and the 4$\sigma$ levels of BF and
CO, positive for H$_2$O, NH$_3$, CH$_4$ and N$_2$, and largest for the $\sigma$
lone pairs of BF and CO (5$\sigma$, $+0.69$ and $+0.83$~eV).

The last three rows of Table~\ref{tab:stats} show its effect. It lowers the mean signed error of
$G_0W_0$@HF from $+0.27$ to $+0.10$~eV and of variant (b) from $+0.35$ to $+0.18$~eV, but
hardly changes their mean absolute errors (0.31 to 0.28 and 0.39 to 0.41~eV), and it makes
variant (a) worse (MSE $-0.40$ to $-0.56$~eV). The static term therefore shifts the levels but
does not change the balance between the screening and the self-energy vertex found above. It is
also sensitive to the basis: in 6-31+G* and aug-cc-pVDZ, with diffuse functions, it is three
to four times larger for the outer-valence levels of Ne and HF (Supplementary Material, Tables S1
and S2).

\subsection{Comparison with QUESTDB}\label{sec:questdb}

For H$_2$O we also compared the STO-3G, DZ [4s2p/2s]\cite{dunning1970} and cc-pVDZ bases
(Supplementary Material, Table~S4). The signs of the errors are the same in all three, but the
ordering of the variants for individual levels changes, and the magnitudes differ by up to
0.3--0.4~eV for (c), so the relative accuracy of the approximations is not basis-independent.
These small bases are not a convergence sequence; the basis-set dependence is assessed against
the QUESTDB best estimates below.

For an external reference, we also evaluated the self-energies in the 6-31+G* and aug-cc-pVDZ
bases at the QUESTDB geometries and compared them with the theoretical best estimates (TBE) of
QUESTDB.\cite{marie2024} The 21 IPs are all valence IPs that QUESTDB lists for the eight
molecules of the main set; the set is limited to these small molecules by the cost of the complete
diagonalization of $\Hbar$, not by a selection among their states. Our $G_0W_0$@HF values
reproduce the QUESTDB $G_0W_0$ values to 0.01~eV. Table~\ref{tab:questdb} summarizes the errors
(individual values in the Supplementary Material, Tables S1 and S2). In aug-cc-pVDZ the mean
absolute errors are 0.36~eV for $G_0W_0$@HF, 0.45~eV for variant (b), 0.74~eV for variant (a) and
1.61~eV for variant (c), against 0.21~eV for IP-EOM-CCSD. The mixed approximation therefore does
not improve upon conventional $G_0W_0$ against independent reference values: replacing the RPA
response by EOM-CCSD and adding exchange at one vertex compensate each other to a large extent,
but the combination is not more accurate than $G_0W_0$. The static self-energy lowers the mean
absolute errors to 0.29~eV for $G_0W_0$@HF and 0.40~eV for (b) and leaves (a) at 0.75~eV. The
mean signed and mean absolute errors in 6-31+G* differ from the aug-cc-pVDZ ones by at most
0.07~eV. The signs of the errors are the same as against IP-EOM-CCSD in cc-pVDZ
(Table~\ref{tab:stats}): (a) underbinds by 0.7~eV on average, (b) and $G_0W_0$@HF overbind by
0.2~eV, and (c) underbinds by 1.5~eV.

\begin{table}
\caption{Errors (eV) relative to the QUESTDB theoretical best estimates for the 21 valence IPs of the eight molecules (individual values in the Supplementary Material, Tables S1 and S2). MSE: mean signed error; MAE: mean absolute error; Max: maximum absolute error.}
\label{tab:questdb}
\begin{ruledtabular}
\begin{tabular}{lrrrrrr}
 & \multicolumn{3}{c}{6-31+G*} & \multicolumn{3}{c}{aug-cc-pVDZ} \\
\cline{2-4}\cline{5-7}
Method & MSE & MAE & Max & MSE & MAE & Max \\
\hline
QUESTDB IP-EOM-CCSD & +0.04 & 0.22 & 1.06 & +0.04 & 0.21 & 1.04 \\
QUESTDB GF2 & $-$0.71 & 0.86 & 1.72 & $-$0.72 & 0.86 & 1.62 \\
$G_0W_0$@HF & +0.18 & 0.42 & 1.56 & +0.19 & 0.36 & 1.46 \\
(a) & $-$0.71 & 0.73 & 2.33 & $-$0.74 & 0.74 & 2.12 \\
(b) & +0.17 & 0.42 & 1.49 & +0.18 & 0.45 & 2.14 \\
(c) & $-$1.44 & 1.56 & 3.00 & $-$1.51 & 1.61 & 3.04 \\
$G_0W_0$@HF + $\Sigma^{(\infty)}$ & +0.17 & 0.29 & 1.44 & +0.18 & 0.29 & 1.45 \\
(a) + $\Sigma^{(\infty)}$ & $-$0.72 & 0.72 & 1.63 & $-$0.75 & 0.75 & 1.61 \\
(b) + $\Sigma^{(\infty)}$ & +0.16 & 0.36 & 1.37 & +0.17 & 0.40 & 2.13 \\
\end{tabular}
\end{ruledtabular}
\end{table}

\section{Conclusions}\label{sec:concl}
Starting from the response theory of NMR spin--spin coupling constants, in which two one-body
perturbations are joined by the electronic response, we have constructed a frequency-dependent
one-particle self-energy by coupling the reference propagator to the neutral EOM-CCSD response
through two interaction vertices, evaluated as a sum over the eigenstates of the
similarity-transformed Hamiltonian without frequency integration or analytic continuation.
Direct, mixed and fully antisymmetrized external vertices define three approximations that
contain GF2 (mixed and antisymmetrized forms; ring part for the direct form), $G_0W_0$@HF and
$G_0W_0$+SOSEX@HF as exact limits for bare and direct-RPA states, so that the screening and the
self-energy vertex can be changed independently within one framework.

For the principal IPs of eight first-row systems, replacing the RPA response by EOM-CCSD lowers
the IPs by 0.45--0.67~eV, whereas exchange at one external vertex raises them by
0.53--0.75~eV. These opposing effects compensate to a large extent, consistent with the findings
of Lewis and Berkelbach\cite{lewis2019} and of Marie and Loos,\cite{marie2025} but the
compensation does not make the mixed approximation more accurate than $G_0W_0$: against the
QUESTDB best estimates\cite{marie2024} in aug-cc-pVDZ the mean absolute errors are 0.45~eV for
the mixed form and 0.36~eV for $G_0W_0$@HF (0.21~eV for IP-EOM-CCSD). The static self-energy
built from the CCSD density, which supplies the static term missing from all three forms, shifts
the levels but does not change this balance.

The third-order analysis shows why. All three forms contain the particle--hole diagrams of the
CC response, but with weights set by the external vertices: the direct form keeps the
polarizability diagrams, the mixed form adds the mixed exchange diagrams of one orientation, and
the fully antisymmetrized form recovers GF2 at second order but assigns weight $\tfrac12$ to every
particle--hole diagram at third order. None contains the explicit particle--particle and
hole--hole ladders or the static term, which at third order are as large as the particle--hole
terms and of opposite sign. Adding the missing ladders alone would therefore not repair the fully
antisymmetrized form; an accurate self-energy requires a balanced treatment of screening, vertex
corrections and all scattering channels, as in Dyson-ADC(3) through third order. Extending the
construction to the particle--particle channel and to larger systems, which requires the
resolvent of $\Hbar$ without its full spectrum, is left for future work.

\section*{Supplementary Material}
See the supplementary material for the comparison with the QUESTDB valence ionization potentials
in the 6-31+G* and aug-cc-pVDZ bases (Tables S1 and S2), the molecular geometries, the basis-set
comparison for H$_2$O, and the numerical checks of the implementation (limiting cases, scaling of
the fluctuation potential, and the third-order expressions against full configuration interaction
and Dyson-ADC(3)).

\begin{acknowledgments}
The authors acknowledge UFIT Research Computing for providing computation resources and
support that have contributed to the research results reported in this publication.
URL: \url{https://it.ufl.edu/rc}
\end{acknowledgments}

\section*{Data availability}
The data that support the findings of this study are available within the article and its
supplementary material. Additional data are available from the corresponding author upon
reasonable request.

\appendix

\section{Perturbative EOM-CCSD response}\label{app:resp}

\emph{Partitioning.} In normal-ordered form with respect to the Hartree--Fock determinant,
$H=E_\mathrm{HF}+F_N+V_N$ with $F_N=\sum_p\epsilon_p\{a^\dagger_pa_p\}$ for canonical orbitals and
$V_N=\tfrac14\sum\braket{pq||rs}\{a^\dagger_pa^\dagger_qa_sa_r\}$, so the fluctuation potential of
Eq.~\eqref{eq:Hlambda} is $W=V_N$ up to a constant, and
$H(\lambda)=E_\mathrm{HF}(\lambda)+F_N+\lambda V_N$. The constant does not enter the self-energy. CC
codes work with $F_N$ and $V_N$ directly: the orbital energies are taken from the converged SCF and
the antisymmetrized integrals from the transformed two-electron file. Scaling the atomic-orbital
two-electron integrals by $\lambda$ after the SCF step and before the transformation therefore
gives exactly $F_N+\lambda V_N$, provided no later step rebuilds the Fock matrix from the scaled
integrals. Written with
operators that are not normal-ordered, the same Hamiltonian has the one-body part
$\epsilon_p\delta_{pq}-\lambda\sum_i\braket{pi||qi}$ and the two-body part $\lambda V$, the
one-body term removing the mean field of the scaled interaction; this form is used in the full
configuration-interaction calculations.

\emph{Amplitudes and Jacobian.} The CCSD amplitudes are
$t^{ab}_{ij}=\lambda\braket{ab||ij}/D_{ijab}+O(\lambda^2)$ with
$D_{ijab}=\epsilon_i+\epsilon_j-\epsilon_a-\epsilon_b$ and $t^a_i=O(\lambda^2)$; the first-order
$\Lambda$ equations, $\lambda^{ij}_{ab}D_{ijab}=\braket{ij||ab}$, give
$\lambda^{ij}_{ab}=t^{ab}_{ij}+O(\lambda^2)$. The EOM-CCSD Jacobian in the space of single and
double excitations is diagonal at zeroth order,
$J^{(0)}_{jb,kc}=\delta_{jk}\delta_{bc}(\epsilon_b-\epsilon_j)$ and similarly for the doubles,
and at first order
\begin{equation}
J^{(1)}_{jb,kc}=\braket{\Phi^b_j|V_N|\Phi^c_k}=\braket{bk||jc},\qquad J^{(1)}_{\mathrm{SD}},\;J^{(1)}_{\mathrm{DS}}\neq0;
\end{equation}
$[F_N,T_2]$ does not contribute to the singles block. The reference component of $R_n$ is
$r_{0,n}=O(\lambda^2)$.

\emph{Transition moments.} For one-body operators $X=\sum_{rs}X_{rs}E_{rs}$ and
$Y=\sum_{rs}Y_{rs}E_{rs}$, with $\tau_{jb}=a^\dagger_ba_j$ and
$\ket{\Phi^{bc}_{jk}}=a^\dagger_ba^\dagger_ca_ka_j\ket{0}$, the moments of Eq.~\eqref{eq:resolvent} are
\begin{align}
l_X(jb)&=\bra{0}(1+\Lambda)\bar X\tau_{jb}\ket{0}\nonumber\\
&=X_{jb}+\sum_{kc}X_{ck}\bra{0}\Lambda_2a^\dagger_ca_ka^\dagger_ba_j\ket{0}+O(\lambda^2)
=X_{jb}+\sum_{kc}X_{ck}\lambda^{jk}_{bc}+O(\lambda^2),\label{eq:l1}\\
r_Y(jb)&=\bra{\Phi^b_j}\bar Y\ket{0}\nonumber\\
&=Y_{bj}+\sum_{kc}Y_{kc}\bra{\Phi^b_j}a^\dagger_ka_cT_2\ket{0}+O(\lambda^2)
=Y_{bj}+\sum_{kc}Y_{kc}t^{bc}_{jk}+O(\lambda^2).\label{eq:r1}
\end{align}
In $l_X$ only the excitation part $X_{ck}a^\dagger_ca_k$ of $X$ can lead, after $\tau_{jb}$, to a
double excitation that $\Lambda_2$ closes; the term $\bra{0}[X,T_2]\tau_{jb}\ket{0}$ vanishes
because $\bra{0}T_2=0$ and $X$ cannot de-excite a triple. In $r_Y$ only the de-excitation part
$Y_{kc}a^\dagger_ka_c$ of $Y$ can bring $T_2\ket{0}$ back to a single excitation, and
$\bra{\Phi^b_j}T_2Y\ket{0}=0$. No normalization enters at this order: the first-order mixing of the
EOM states is in $\bm{J}^{(1)}$, inside the resolvent, and $r_{0,n}$ is of second order. In the
commutator form $\bra{0}(1+\Lambda)[\bar X,\tau_{jb}]\ket{0}$ the subtracted term
$\bra{0}\Lambda_2\tau_{jb}X\ket{0}=\sum_{kc}X_{ck}\lambda^{jk}_{bc}$ cancels the first-order term of
Eq.~\eqref{eq:l1}, so that form has no first-order correction to the right moment and loses
the $\Lambda_2$ term, which carries half of the ground-state-correlation correction at third
order (Appendix~\ref{app:sigma3}). The doubles
components $l_X(\mathrm{D})$ and $r_Y(\mathrm{D})$ are of first order, but the zeroth-order moments
of a double excitation vanish, so every path through the doubles block contains two first-order
factors. The first-order response is therefore Eq.~\eqref{eq:chi1} with singles only.

\emph{Forward and backward parts.} The time-ordered response of Eq.~\eqref{eq:chiXY} has a
forward part, with poles at $z=\omega_n-i0^+$, and a backward part, $-Y^R_nX^L_n/(z+\omega_n-i0^+)$,
with poles at $z=-\omega_n+i0^+$. The self-energy is the convolution
$\Sigma_{pq}(\omega)=\kappa\,(i/2\pi)\int d\omega'\,e^{i\omega'0^+}\sum_mG^0_m(\omega+\omega')\,
\langle\!\langle h^{(mp)};h^{(qm)}\rangle\!\rangle_{\omega'}$, closed in the upper half plane. For
$m=i$ occupied the pole of $G^0_i$ lies in the upper half plane; the forward part, whose poles lie
in the lower half plane, gives the residue at the pole of $G^0_i$, while the backward part has both
its poles in the upper half plane and gives zero. For $m=a$ virtual the pole of $G^0_a$ lies in the
lower half plane; only the backward part has a pole inside the contour. The hole ordering
therefore takes the forward part of the response, and the particle ordering takes its backward
part. Since the backward part of $\langle\!\langle X;Y\rangle\!\rangle$ at $z$ equals the forward part
of $\langle\!\langle Y;X\rangle\!\rangle$ at $-z$, and this gives the second term of
Eq.~\eqref{eq:sigma_chi}, with the two operators in exchanged order. Both parts of the response
are therefore used, each once, and the perturbative expansion of the backward part is
Eq.~\eqref{eq:chi1} with $X$ and $Y$ exchanged. The overall sign and $\kappa$ are fixed by the
bare-state limit.

\section{Third-order self-energy}\label{app:sigma3}

\emph{Exact third order.} All sums below are over unrestricted spin-orbital indices, and
$D^{P}_{abj}=\omega+\epsilon_j-\epsilon_a-\epsilon_b$, $D^{H}_{ija}=\omega+\epsilon_a-\epsilon_i-\epsilon_j$.
The exact third-order self-energy of $H(\lambda)$ is the third-order part of the Dyson
ADC(3) self-energy of Schirmer, Cederbaum and Walter,\cite{schirmer1983,vonniessen1984}
$\Sigma(\omega)=\Sigma(\infty)+\bm{U}^\dagger(\omega-\bm{K}-\bm{C})^{-1}\bm{U}$, with
$\bm{U}=\bm{U}^{(1)}+\bm{U}^{(2)}$ and $\bm{C}=\bm{C}^{(1)}$ in the space of 2p1h and 2h1p
configurations. Expanding to third order [Eq.~(43) of Ref.~\onlinecite{schirmer1983}],
\begin{equation}
\Sigma^{(3)}=\bm{U}^{(1)\dagger}\bm{R}\,\bm{C}^{(1)}\bm{R}\,\bm{U}^{(1)}
+\bigl[\bm{U}^{(1)\dagger}\bm{R}\,\bm{U}^{(2)}+\bm{U}^{(2)\dagger}\bm{R}\,\bm{U}^{(1)}\bigr]
+\Sigma^{(3)}(\infty)
\equiv\Sigma^{(3)}_C+\Sigma^{(3)}_U+\Sigma^{(3)}_\infty,
\label{eq:adc3}
\end{equation}
where $\bm{R}=(\omega-\bm{K})^{-1}$.
Written out with the first-order couplings, the first-order coupling matrix and the second-order
coupling amplitudes of Ref.~\onlinecite{schirmer1983} [its Eqs.~(40a), (42c), (44b) and (45)],
the first-order coupling term is
\begin{align}
\Sigma^{(3)}_{C,pp}={}&\tfrac14\sum\frac{\braket{ab||pj}\braket{ab||cd}\braket{cd||pj}}{D^P_{abj}D^P_{cdj}}
+\sum\frac{\braket{ab||pj}\braket{bk||jd}\braket{ad||pk}}{D^P_{abj}D^P_{adk}}\nonumber\\
&-\tfrac14\sum\frac{\braket{ij||pa}\braket{kl||ij}\braket{kl||pa}}{D^H_{ija}D^H_{kla}}
+\sum\frac{\braket{ij||pa}\braket{bj||al}\braket{il||pb}}{D^H_{ija}D^H_{ilb}},
\label{eq:sigC}
\end{align}
which follows from the coupling matrix of the 2p1h configurations
$a^\dagger_aa^\dagger_ba_j\ket{0}$,
$C_{abj,cdk}=\delta_{jk}\braket{ab||cd}+\mathcal{P}(ab)\mathcal{P}(cd)\,\delta_{ac}\braket{bk||jd}$, and
of the 2h1p configurations $a^\dagger_aa_ja_i\ket{0}$,
$C_{ija,klb}=\delta_{ab}\braket{kl||ij}-\mathcal{P}(ij)\mathcal{P}(kl)\,\delta_{ik}\braket{bj||al}$,
with couplings $\braket{ab||pj}$ and $\braket{ij||pa}$ to the one-particle line, as in the
algebraic diagrammatic construction.\cite{schirmer1983} The first and third terms of
Eq.~\eqref{eq:sigC} are the particle--particle and hole--hole ladders, Fig.~\ref{fig:pt3}(c); the
second and fourth are the particle--hole terms, Fig.~\ref{fig:pt3}(a). With the second-order
coupling amplitudes $\bm{U}^{(2)}$, the coupling corrections are
\begin{align}
\Sigma^{(3)}_{U,pp}={}&2\sum\frac{\braket{ab||pj}}{D^P_{abj}}\sum_{kc}\braket{ak||pc}t^{bc}_{jk}
+2\sum\frac{\braket{ij||pa}}{D^H_{ija}}\sum_{kc}\braket{pk||ic}t^{ac}_{jk}\nonumber\\
&+\sum\frac{\braket{ab||pj}}{D^P_{abj}}\tfrac12\sum_{kl}\braket{kl||pj}t^{ab}_{kl}
+\sum\frac{\braket{ij||pa}}{D^H_{ija}}\tfrac12\sum_{cd}\braket{cd||pa}t^{cd}_{ij},
\label{eq:sigU}
\end{align}
of particle--hole type [first line, Fig.~\ref{fig:pt3}(b)] and of ladder type [second line,
Fig.~\ref{fig:pt3}(d)]. The static part is given exactly by the correlated one-body density,
$\Sigma_{pq}(\infty)=\sum_{rs}\braket{pr||qs}(\rho-\rho^\mathrm{HF})_{rs}$ [Eq.~(25) of
Ref.~\onlinecite{schirmer1983}], so at third order
\begin{equation}
\Sigma^{(3)}_{\infty,pp}=\sum_{rs}\braket{pr||ps}\,\rho^{(2)}_{rs},
\label{eq:sigS}
\end{equation}
with the second-order density $\rho^{(2)}_{ij}=-\tfrac12\sum_{kab}t^{ab}_{ik}t^{ab}_{jk}$,
$\rho^{(2)}_{ab}=\tfrac12\sum_{ijc}t^{ac}_{ij}t^{bc}_{ij}$ and
$\rho^{(2)}_{ia}=\rho^{(2)}_{ai}=t^{a\,(2)}_i=\bigl(\tfrac12\sum_{jbc}\braket{aj||bc}t^{bc}_{ij}
-\tfrac12\sum_{jkb}\braket{jk||ib}t^{ab}_{jk}\bigr)/(\epsilon_i-\epsilon_a)$
[Fig.~\ref{fig:pt3}(e)].

\emph{Variants.} Inserting Eq.~\eqref{eq:chi1} into Eq.~\eqref{eq:sigma_chi} gives the
third-order term of each variant,
\begin{align}
\Sigma^{(3)}_{pq}={}&-\kappa\sum_i\Bigl[\bm{l}^{(0)}\bm{R}_0\bm{J}^{(1)}\bm{R}_0\bm{r}^{(0)}
+\bm{l}^{(1)}\bm{R}_0\bm{r}^{(0)}+\bm{l}^{(0)}\bm{R}_0\bm{r}^{(1)}\Bigr]_{X=h^{(ip)}_P,\,Y=h^{(qi)}_Q,\,z=\epsilon_i-\omega}\nonumber\\
&+\kappa\sum_a\Bigl[\cdots\Bigr]_{X=h^{(qa)}_Q,\,Y=h^{(ap)}_P,\,z=\omega-\epsilon_a},
\end{align}
where $h_P$, $h_Q$ are the operator of Eq.~\eqref{eq:hpm} or its direct part, according to the
vertex. With antisymmetrized vertices on both sides each term is, configuration by
configuration, one of the particle--hole terms of Eqs.~\eqref{eq:sigC} and~\eqref{eq:sigU}: the
$\bm{J}^{(1)}$ term reproduces the second and fourth terms of Eq.~\eqref{eq:sigC}, and the
$\bm{l}^{(1)}$ and $\bm{r}^{(1)}$ terms each reproduce one half of the first line of
Eq.~\eqref{eq:sigU}. A 2p1h configuration $(a,b,j)$ is reached from $m=a$ with the pair $(j,b)$ and
from $m=b$ with the pair $(j,a)$, but the particle--hole interaction acts only within the pair, so
each term of Eq.~\eqref{eq:sigC} with a shared particle is generated once; the factor
$\kappa=\tfrac12$, which compensates the two orientations at second order, then halves it. The
ladders and the ladder-type corrections need an interaction between the intermediate line and the
pair, or a ground-state pair attached to the intermediate line, and do not occur.

For the split at the outer vertices, the couplings are written as
$\braket{ab||pj}=\braket{ab|pj}-\braket{ab|jp}$ with $a$ the intermediate line, i.e.\ as a direct
($d$) coupling $(ap|bj)$ and an exchanged ($x$) coupling $(aj|bp)$, and likewise $\braket{ij||pa}$
and the integrals $\braket{ak||pc}$, $\braket{pk||ic}$ in the coupling corrections. Each
particle--hole term of Eqs.~\eqref{eq:sigC} and~\eqref{eq:sigU} is bilinear in the two outer
couplings and splits into four parts with unit weight, according to whether the attachment at the
$p$ vertex and at the $q$ vertex is direct or exchanged; for the coupling corrections, the
integral of the correction counts toward the vertex at which it sits. Variant (a) keeps the part
that is direct at both vertices, (b) the two parts that are direct at $p$, and (c) all four with
weight $\tfrac12$ (Table~\ref{tab:pt3}).

\emph{Worked example.} We use real spin orbitals, canonical orbital energies, unrestricted sums,
and the diagonal element $p=q$; the off-diagonal case is treated below. Take the $\bm{J}^{(1)}$
term in the particle ordering. There
$X=h^{(pa)}_Q$ and $Y=h^{(ap)}_P$ (Eq.~\eqref{eq:sigma_chi}), $z=\omega-\epsilon_a$, and the
zeroth-order moments are the single elements $l^{(0)}_X(jb)=X_{jb}$ and
$r^{(0)}_Y(kc)=Y_{ck}$; for antisymmetrized vertices $X_{jb}=\braket{pj||ab}$ and
$Y_{ck}=\braket{ac||pk}$, for direct ones $\braket{pj|ab}$ and $\braket{ac|pk}$. With
$[\bm{R}_0]_{jb}=(\omega-\epsilon_a-\epsilon_b+\epsilon_j)^{-1}=1/D^P_{abj}$,
\begin{equation}
\Sigma^{(3)}_{J,P}=\kappa\sum_a\sum_{jb}\sum_{kc}\frac{X_{jb}\,\braket{bk||jc}\,Y_{ck}}{D^P_{abj}\,D^P_{ack}} .
\end{equation}
The particle $a$ runs from the lower ($p$) to the upper ($q$) outer vertex; the pair $k\to c$ is
created at the lower vertex by $Y$, scattered into $j\to b$ by $\braket{bk||jc}$, and removed at
the upper vertex by $X$.
This is Fig.~\ref{fig:pt3}(a). For variant (c), $\kappa=\tfrac12$ and
$\Sigma^{(3)}_{J,P}=\tfrac12\sum_{abcjk}\braket{ab||pj}\braket{bk||jc}\braket{ac||pk}/(D^P_{abj}D^P_{ack})$,
which is $\tfrac12$ times the second term of Eq.~\eqref{eq:sigC}. That term has weight 1 because the
four $\delta$ terms of $C_{abj,cdk}$ are equivalent and cancel the $\tfrac14$ from unrestricted
sums. In the variant the intermediate line is labelled, and a pair of configurations
$(a,b,j)$, $(a,c,k)$ that share the particle $a$ is connected by $\bm{J}^{(1)}$ only in the
orientation $m=a$; the orientation $m=b$ that doubles the second-order term has no partner,
because $(a,c,k)$ does not contain $b$. So the second-order factor $\kappa=\tfrac12$ is not
compensated, and the weight is $\tfrac12$. For variant (a), with $\kappa=1$ and both elements
direct, the expression is
$\sum\braket{ab|pj}\braket{bk||jc}\braket{ac|pk}/(D^P_{abj}D^P_{ack})$, which is the $dd$ part of the
same exact term [Fig.~\ref{fig:pt3}(f)] with weight 1. For variant (b) the $q$ element is
antisymmetrized, $\braket{pj||ab}=\braket{pj|ab}-\braket{pj|ba}$, which adds the mixed part with the
exchange at the $q$ vertex, which in this ordering is the upper one [Fig.~\ref{fig:pt3}(g)], with weight 1, and contains neither the part exchanged at the $p$
vertex nor $xx$.

The ground-state terms work the same way. With Eq.~\eqref{eq:r1}, the $\bm{r}^{(1)}$ term in the
particle ordering is
$\kappa\sum_{ajb}X_{jb}\bigl(\sum_{kc}Y_{kc}t^{bc}_{jk}\bigr)/D^P_{abj}$, with
$Y_{kc}=\braket{ak||pc}$; with Eq.~\eqref{eq:l1}, the $\bm{l}^{(1)}$ term is
$\kappa\sum_{ajb}\bigl(\sum_{kc}X_{ck}\lambda^{jk}_{bc}\bigr)Y_{bj}/D^P_{abj}$, with
$X_{ck}=\braket{pc||ak}$. For $p=q$ and real orbitals the two are equal, and each is $\kappa/2$
times the first term of Eq.~\eqref{eq:sigU}, whose factor 2 counts the correction at either outer
vertex. In variant (c) they add to $\tfrac12$ of that term; with the commutator form of the right
moment, the $\bm{l}^{(1)}$ term would be missing and (c) would contain $\tfrac14$. The hole
ordering is analogous, with $\braket{bj||al}$, $\braket{pk||ic}$ and $D^H$.

Equations~\eqref{eq:sigC} and~\eqref{eq:sigU} are the corresponding terms of
Eq.~\eqref{eq:adc3} summed over unrestricted indices; evaluating both forms for random
antisymmetrized interaction tensors gives identical diagonal and off-diagonal elements. The
expressions, the weights of Table~\ref{tab:pt3} and the implementation were also checked against
full configuration interaction (Supplementary Material).

\begin{table}
\caption{Third-order contributions to $\Sigma_{pp}(\epsilon_p)$ (eV) for H$_2$O in the STO-3G
basis: exact parts [Eqs.~\eqref{eq:sigC}--\eqref{eq:sigS}] and the three variants.}
\label{tab:mag}
\begin{ruledtabular}
\begin{tabular}{lrrrrrrrrr}
MO & ph, $C$ & ph, $U$ & pp/hh, $C$ & pp/hh, $U$ & static & exact & (a) & (b) & (c) \\
\hline
2a$_1$ & 10.00 & $-0.35$ & $-7.46$ & 0.09 & $-0.08$ & 2.20 & 7.58 & 3.62 & 4.83 \\
1b$_2$ & 0.29 & $-0.42$ & $-0.23$ & 0.22 & 0.01 & $-0.14$ & 0.12 & 0.01 & $-0.07$ \\
3a$_1$ & 0.97 & $-0.53$ & $-0.76$ & 0.15 & $-0.07$ & $-0.24$ & $-0.06$ & $-0.17$ & 0.22 \\
1b$_1$ & 1.54 & $-0.57$ & $-1.23$ & 0.03 & $-0.17$ & $-0.38$ & $-0.19$ & $-0.24$ & 0.49 \\
\end{tabular}
\end{ruledtabular}
\end{table}

\bibliography{refs}

\end{document}